\documentclass[preprint,12pt]{amsart}
\usepackage{comment}

\usepackage[english]{babel}
\usepackage[T1]{fontenc}
\usepackage[utf8]{inputenc}
\usepackage{amsmath,amssymb,mathrsfs}
\usepackage{amscd}
\usepackage{tikz}
\numberwithin{equation}{section}
\newtheorem{theo}{Theorem}[section]
\newtheorem{cor}{Corollary}[section]
\newtheorem{prop}{Proposition}[section]
\newtheorem{lem}{Lemma}[section]

\numberwithin{figure}{section}

\newenvironment{prof}
	{\textit{\textbf{Proof.}}}
	{\hfill $\blacksquare$\vskip 8pt}

\title[Lieb-Thirring type inequalities on complex eigenvalues]{Lieb-Thirring type inequalities for non self-adjoint perturbations of magnetic Schrödinger operators}
\author{Diomba \textsc{Sambou}}
\address{Institut de Mathématiques de Bordeaux, Université Bordeaux 1, 351 cours de la Libération F-33405 Talence cedex}
\email{diomba.sambou@math.u-bordeaux1.fr}

\keywords{Magnetic Schrödinger operators, Lieb-Thirring type inequalities, non self-adjoint relatively compact perturbations.}
\subjclass[2010]{Primary: 35P20; Secondary: 47A75, 47A55.}

\begin{document}

\begin{abstract}
Let $H := H_{0} + V$ and $H_{\perp} := H_{0,\perp} + V$ be respectively perturbations of the free Schrödinger operators $H_{0}$ on $L^{2}\big(\mathbb{R}^{2d+1}\big)$ and $H_{0,\perp}$ on $L^{2}\big(\mathbb{R}^{2d}\big)$, $d \geq 1$ with constant magnetic field of strength $b>0$, and $V$ is a complex relatively compact perturbation. We prove Lieb-Thirring type inequalities for the discrete spectrum of $H$ and $H_{\perp}$. In particular, these estimates give $a\, priori$ information on the distribution of the discrete eigenvalues around the Landau levels of the operator, and describe how fast sequences of eigenvalues converge.
\end{abstract}

\maketitle

\section{Introduction}

Let $\textup{\textbf{x}} := (X_{\perp},x) \in \mathbb{R}^{2d + 1}$ be the Cartesian coordinates, with $d \geq 1$ and $X_{\perp} := (x_{1},y_{1},\ldots, x_{d},y_{d}) \in \mathbb{R}^{2d}$. Let $b > 0$ be a constant which is the strength of the magnetic field and consider
\begin{equation}\label{eq0,1}
\displaystyle H_{0,\perp} := \sum_{j=1}^{d} \left\lbrace \left( D_{x_{j}} + \frac{1}{2}by_{j} \right)^{2} + \left( D_{y_{j}} - \frac{1}{2}bx_{j} \right)^{2} \right\rbrace,
\end{equation}
\begin{equation}\label{eq0,2}
\displaystyle H_{0} := H_{0,\perp} + D_{x}^{2}, \hspace{0.5cm} D_{\nu} := -i\frac{\partial}{\partial \nu},
\end{equation}
the Schrödinger operators with constant magnetic field. The self-adjoint operators $H_{0,\perp}$ and $H_{0}$ are originally defined on $C_{0}^{\infty} \big( \mathbb{R}^{n} \big)$ for $n = 2d$ and $n = 2d + 1$ respectively, and then closed in $L^{2} \big( \mathbb{R}^{n} \big)$. It is well known \big(see e.g. \cite{dim}\big) that the spectrum of the operator $H_{0,\perp}$ consists of the increasing sequence of Landau levels $\Lambda_{j}$, $j \in \mathbb{N} := \lbrace 0,1,2 \ldots \rbrace$:
\begin{equation}\label{eq1,00}
\Lambda_{j} = b(d + 2j), \quad j \in \mathbb{N} := \lbrace 0,1,2 \ldots \rbrace,
\end{equation}
and the multiplicity of each eigenvalue $\Lambda_{j}$ is infinite. In the sequel we put $E := \lbrace \Lambda_{j} \rbrace_{j \in \mathbb{N}}$ for the set of Landau levels. Since the operator $H_{0}$ can be written in $L^{2} \big( \mathbb{R}^{n} \big) = L^{2} \big( \mathbb{R}^{2d} \big) \otimes L^{2} \big( \mathbb{R} \big)$ as $$H_{0} = H_{0,\perp} \otimes I + I \otimes D_{x}^{2},$$ the spectrum of $H_{0}$ is absolutely continuous, equals $[\Lambda_{0},+\infty)$, and has an infinite set of thresholds $\Lambda_{j}$, $j \in \mathbb{N}$.

On the domains of $H_{0}$ and $H_{0,\perp}$, we introduce the perturbed operators 
\begin{equation}\label{eq0.3}
H := H_{0} + V \quad \text{and} \quad H_{\perp} := H_{0,\perp} + V,
\end{equation}
where  $V : \mathbb{R}^{n} \rightarrow \mathbb{C}$ is a non self-adjoint perturbation for $n = 2d + 1$ and $n = 2d$ respectively. Everywhere in this article, we identify $V$ with the multiplication operator by the function $V$. To simplify in what follows below, by $\mathcal{H}_{0}$ we mean the free operators $H_{0}$ and $H_{0,\perp}$, and by $\mathcal{H}$ we mean $H$ and $H_{\perp}$ defined by \eqref{eq0.3}. So let 
$$N(\mathcal{H}) := \lbrace (\mathcal{H}f,f):f \in \text{dom}(\mathcal{H}), \Vert f \Vert_{L^{2}} = 1 \rbrace$$ 
be the numerical range of $\mathcal{H}$. It is well known \big(see e.g. \cite[Lemma 9.3.14]{dav}\big) that the spectrum $\sigma(\mathcal{H})$ of $\mathcal{H}$ satisfies $$\sigma(\mathcal{H}) \subseteq \overline{N(\mathcal{H})}.$$ Thus, if the perturbation $V$ is bounded we can easily verify that 
\begin{equation}\label{eq0.03}
\sigma(\mathcal{H}) \subseteq \overline{N(\mathcal{H})} \subset \big\lbrace \lambda \in \mathbb{C} : \hspace{0.1cm}\textup{Re} \hspace{0.3mm} \lambda \geq -\Vert V \Vert_{\infty} \hspace{0.2cm} \textup{and} \hspace{0.2cm} \vert \textup{Im} \hspace{0.3mm} \lambda \vert \leq \Vert V \Vert_{\infty}  \big\rbrace.
\end{equation}

Now let us introduce some conventional definitions we will use in this paper. Let $Z$ be a closed operator acting on a separable Hilbert space $X$, and denote by $R(z)$ its resolvent. If $\lambda$ is an isolated point of the spectrum $\sigma(Z)$ of $Z$, recall that the Riesz projection of $Z$ with respect to $\lambda$ is defined by
$$P_{\lambda} := \frac{1}{2i\pi} \int_{\gamma} R(z)dz,$$
where $\gamma$ is a small positively oriented circle centered at $\lambda$, and containing $\lambda$ as the only point of $\sigma(Z)$. We shall say that $\lambda$ is \textit{a discrete eigenvalue} of $Z$ if its algebraic multiplicity
$$m(\lambda) := \text{rank} \big( P_{\lambda} \big)$$
is finite. The algebraic multiplicity of $\lambda$ can be bigger than its geometric multiplicity defined as the dimension of the eigenspace $\text{ker}(Z - \lambda)$. In the sequel, the set of discrete eigenvalues of $Z$ is called the \textit{discrete spectrum} and is denoted by $\sigma_{d}(Z)$. The \textit{essential spectrum} $\sigma_{\textup{ess}}(Z)$ of $Z$ is defined as the set of complex numbers $\lambda$ such that $Z - \lambda$ is not a Fredholm operator. It's a closed subset of $\sigma(Z)$. We denote by $S_{q}(X)$, $q \in [1,+\infty)$ the Schatten-von Neuman classes of compact linear operators $L$ for which the norm $\Vert L \Vert_{q} := (\textup{Tr} \hspace{0.5mm} \vert L \vert^{q})^{1/q}$ is finite. 

Assume that $V$ is relatively compact with respect to $\mathcal{H}_{0}$. Since $\sigma \big( \mathcal{H}_{0} \big) \subseteq \mathbb{R}$, then it follows from a version of the Weyl criterion \cite[Corollary 2.3.3]{demu} that $\sigma_{\textup{ess}} (H) = [\Lambda_{0},+\infty)$ and $\sigma_{\textup{ess}} (H_{\perp}) = \lbrace \Lambda_{j} \rbrace_{j \in \mathbb{N}}$. However, the discrete spectrum $\sigma_{d}(\mathcal{H})$ of $\mathcal{H}$ can only accumulate on $\sigma_{\textup{ess}} (\mathcal{H})$. For a separable Hilbert space $X$, we denote by $S_{q}(X)$, $q \in [1,+\infty)$ the Schatten-Von Neuman classes of compact linear operators $L$ for which the norm $\Vert L \Vert_{q} := (\textup{Tr} \hspace{0.5mm} \vert L \vert^{q})^{1/q}$ is finite. Throughout this article, we consider non self-adjoint electric potentials $V$ bounded and satisfying the following estimates:
\begin{equation}\label{eq1,1}
\begin{aligned}
\vert V(\textup{\textbf{x}}) \vert & \leq \hspace{-1cm}     &     & \hspace{-0.2cm} C F(\textup{\textbf{x}}) G(x), \quad \text{if} \quad n = 2d + 1, \\
\vert V(X_{\perp}) \vert & \leq                             &                             & \hspace{-0.2cm} C F(X_{\perp}), \quad \hspace{0.45cm} \text{if} \quad n = 2d,
\end{aligned}
\end{equation}
where $C > 0$ is a constant in both cases, $F$ and $G$ are two positive functions satisfying $F \in L^{p}(\mathbb{R}^{n})$ for $p \geq 2$, and $G \in \big( L^{2} \cap L^{\infty} \big) \big{(} \mathbb{R} \big{)}$. Under these assumptions, we obtain some estimates of the $p$-Schatten norm of the sandwiched resolvents $F (H_{0} - \lambda )^{-1} G$ and $F (H_{0,\perp} - \lambda )^{-1}$ (see Lemma \ref{lem1} and Lemma \ref{lem01}, respectively). Furthermore, we use these estimates to obtain quantitative bounds on  $\sigma_{d}(\mathcal{H})$. In particular, these estimates give $a\, priori$ information on the distribution of eigenvalues around the Landau levels of the operator. Most of known results on $\sigma_{d}(\mathcal{H})$ deal with selfadjoint perturbations $V$ and investigate the asymptotic behaviour of $\sigma_{d}(\mathcal{H})$ near the boundary points of its essential spectrum. This behaviour has been extensively studied in case where $V$ admits power-like or slower decay at infinity \big(see \cite[chap. 11-12]{iv}, \cite{pus}, \cite{ra}, \cite{rai}, \cite{sob}, \cite{tam}\big). In \cite{raik} this behaviour is studied for potentials $V$ decaying at infinity exponentially fast or having a compact support. For Landau Hamiltonians in exterior domains see \cite{kac}, \cite{per} and \cite{push}.

Typical example of potentials satisfying \eqref{eq1,1} for $n = 2d + 1$ is the special case of the relatively compact perturbations $V : \mathbb{R}^{n} \rightarrow \mathbb{C}$ satisfying the estimate
\begin{equation}\label{eq1,0}
\vert V(\textup{\textbf{x}}) \vert \leq C \hspace{0.5mm} \langle X_{\perp} \rangle^{-m_{\perp}} \langle x \rangle^{-m}, \quad m_{\perp} > 0, \quad  m > 1/2,
\end{equation}
where $\langle y \rangle : = \big( 1 + \vert y \vert^{2} \big)^{1/2}$, $y \in \mathbb{R}^{d}$, $d \geq 1$. Indeed, put $F(\textup{\textbf{x}}) =  \langle X_{\perp} \rangle^{-m_{\perp}} \langle x \rangle^{-\nu}$ and $G(x) = \langle x \rangle^{-\tilde{m}}$, where $\nu + \tilde{m} = m$ with $\nu > 0$ and $\tilde{m} > 1/2$. Clearly for any $p \geq 2$ such that $p \hspace*{0.3mm} m_{\perp} > 2d$ and $p \nu > 1$, $F \in L^{p}(\mathbb{R}^{n})$ and $G \in \big( L^{2} \cap L^{\infty} \big) \big{(} \mathbb{R} \big{)}$. We can also consider perturbations $V : \mathbb{R}^{n} \rightarrow \mathbb{C}$ verifying
\begin{equation}\label{eq0,52}
\vert V(\textup{\textbf{x}}) \vert \leq C \hspace{0.5mm} \langle \textup{\textbf{x}} \rangle^{-\alpha}, \quad \alpha > 1/2.
\end{equation}
Indeed, \eqref{eq0,52} implies that 
$$\vert V(\textup{\textbf{x}}) \vert \leq C \hspace{0.5mm} \langle \textup{\textbf{x}} \rangle^{-(\alpha-\nu)} \langle x \rangle^{-\nu}, \quad \nu \in (1/2,\alpha).$$ 
So $V$ satisfies \eqref{eq1,1} for any $p \geq 2$ with $p(\alpha-\nu) > n$. Note that \eqref{eq0,52} implies \eqref{eq1,0} with any $m \in (1/2,\alpha)$ and $m_{\perp} = \alpha - m$.

In the $2d$-dimensional case, assumption \eqref{eq1,1} is satisfied for example by the class of potentials $V : \mathbb{R}^{n} \rightarrow \mathbb{C}$ such that
\begin{equation}\label{eq0,51}
\vert V(X_{\perp}) \vert \leq C \hspace{0.5mm} \langle X_{\perp} \rangle^{-m_{\perp}}, \quad m_{\perp} > 0,
\end{equation}
$p \hspace*{0.3mm} m_{\perp} > 2d$ and $p \geq 2$. Under the assumption $m_{\perp} > 0$, $p \hspace*{0.3mm} m_{\perp} > 2d$ and $p \geq 2$, we can also consider power-like decaying electric potentials $V : \mathbb{R}^{n} \rightarrow \mathbb{C}$ satisfying the asymptotic property
\begin{equation}\label{eq0,5}
V(X_{\perp}) = \text{v}\big(X_{\perp}/\vert X_{\perp} \vert\big) \vert X_{\perp} \vert^{-m_{\perp}} \big( 1 + o(1) \big) \quad \text{as} \quad \vert X_{\perp} \vert \rightarrow \infty,
\end{equation}
where $\text{v}$ is a continuous function on $\mathbb{S}^{2d-1}$ which is non-negative and does not vanish identically \big(see also \cite{ra} where $V$ is self-adjoint\big).

To prove our first result (see Theorem \ref{theo1}), we first construct a holomorphic function whose zeros coincide with the eigenvalues of $H$. Moreover, we use a result by Borichev, Golinskii and Kupin \cite{bor} and complex analysis methods to  get information on these zeros. Similar techniques are used in \cite{dem} and \cite{demu} for non-magnetic Schrödinger operators $H$ with $\sigma_{\textup{ess}} (H) = [0,+\infty)$, and Jacobi matrices $J$ with $\sigma_{\textup{ess}} (J) = [-2,2]$. In both situations, the essential spectrum has a finite number of thresholds ($0$ for the first and $-2$, $2$ for the second). Since in our case $\sigma_{\textup{ess}} (H) = [\Lambda_{0},+\infty)$ with an infinite set of thresholds $\Lambda_{j}$, we are led to introduce appropriate modifications to the above techniques to prove our results. More precisely, we will obtain two types of estimates. First, we bound the sums depending on parts of $\sigma_d(\mathcal{H})$ concentrated around a Landau level (see Proposition \ref{pro1}) using the Schwarz-Christoffel formula. Second, we get global estimates summing up the previous bounds with appropriate weights. This is to compare to results of \cite{dem} and \cite{demu}, where global estimates were obtained directly  by mapping conformally $\overline{\mathbb{C}} \setminus [0,+\infty)$ and $\overline{\mathbb{C}} \setminus [-2,2]$ onto the unit disk respectively \big(see also \cite{bor}\big). To prove our second main result (see Theorem \ref{theo3}), we reason similarly to \cite{gol}; in particular, we use a recent result by Hansmann \cite{han} and a technical distortion lemma (see Lemma \ref{lem02}).

The paper is organized as follows. In section 2, we formulate our main results and we discuss some of their immediate consequences on the discrete spectrum of the operators $H$ and $H_{\perp}$ defined by \eqref{eq0.3}. Sections 3, 4 and 5 are devoted to the $(2d + 1)$-dimensional Schrödinger operators $H_{0}$ and $H$. In section 3, we establish estimates on appropriate sandwiched resolvents. Section 4 contains auxiliary material as the construction of a holomorphic function whose zeros coincide with the eigenvalues of $H$ in $\mathbb{C} \setminus [\Lambda_{0},+\infty)$, and the presentation of appropriate tools of complex analysis. In section 5, we prove a local bound on the eigenvalues of the operator $H$ (see Proposition \ref{pro1}) and derive the proof of Theorem \ref{theo1} from it. Section 6 is devoted to the $2d$-dimensional Schrödinger operators $H_{0,\perp}$ and $H_{\perp}$, and we prove Theorem \ref{theo3}.

\section{Main Results}

In this section and elsewhere in this paper, for any $r \in \mathbb{R}$ we denote by $r_{+} := \max \hspace{0.5mm} (r,0)$ and by $[r]$ its integer part. Otherwise, each $\lambda \in \sigma_{d} (\mathcal{H})$ is considered accordingly to its algebraic multiplicity, where $\mathcal{H}$ is the operator $H$ or $H_{\perp}$ defined by \eqref{eq0.3}.

\subsection{The $(2d + 1)$-dimensional case}

We obtain a Lieb-Thirring type inequality for the discrete spectrum of the $(2d + 1)$-dimensional Schrödinger operator $H$ defined by \eqref{eq0.3}. The following theorem is proved in section 5.

\begin{theo}\label{theo1}
Let $H = H_{0} + V$ with $V$ satisfying \eqref{eq1,1} for $n = 2d + 1$, $d \geq 1$. Assume that $F \in L^{p}\big(\mathbb{R}^{n}\big)$ with $p \geq 2 \big[ \frac{d}{2} \big] + 2$ and $G \in \big( L^{2} \cap L^{\infty} \big) \big{(} \mathbb{R} \big{)}$. Define
\begin{equation}\label{eq1,005}
K := \Vert F \Vert_{L^{p}}^{p} \big{(} \Vert G \Vert_{L^{2}} + \Vert G \Vert_{L^{\infty}} \big{)}^{p} \big{(} 1 + \Vert V \Vert_{\infty} \big{)}^{d  + \frac{p}{2} + \frac{3}{2} + \varepsilon} .
\end{equation} 
for $0 < \varepsilon < 1$. Then we have
\begin{equation}\label{est0}
\displaystyle \sum_{\lambda \hspace{0.5mm} \in \hspace{0.5mm} \sigma_{d}(H)} \frac{\textup{dist} \big( \lambda,[\Lambda_{0},+\infty )\big)^{\frac{p}{2} + 1 + \varepsilon} \hspace{0.5mm} \textup{dist} (\lambda,E)^{(\frac{p}{4} - 1 + \varepsilon)_{+}}}{(1 + \vert \lambda \vert)^{\gamma}} \leq C_{0} K,
\end{equation}
where $E$ is the set of Landau levels defined by \eqref{eq1,00}, $\gamma > d + \frac{3}{2}$ and $C_{0} = C(p,b,d,\varepsilon)$ is a constant depending on $p$, $b$, $d$ and $\varepsilon$.
\end{theo}

Since for any $\tau > 0$ with $\vert \lambda \vert \geq \tau$ we have
\begin{equation}\label{eq84}
\frac{1}{1 + \vert \lambda \vert} = \frac{1}{\vert \lambda \vert} \frac{1}{1 + \vert \lambda \vert^{-1}} \geq \frac{1}{\vert \lambda \vert} \frac{1}{1 + \tau^{-1}},
\end{equation}
the following holds.

\begin{cor}
Under the assumptions and the notations of Theorem $\ref{theo1}$, the following bound holds for any $\tau > 0$
\begin{equation}\label{est1}
\displaystyle \sum_{\substack{\lambda \hspace{0.5mm} \in \hspace{0.5mm} \sigma_{d}(H) \\ \vert \lambda \vert \hspace{0.5mm} \geq \hspace{0.5mm} \tau}} \frac{\textup{dist} \big( \lambda,[\Lambda_{0},+\infty )\big)^{\frac{p}{2} + 1 + \varepsilon} \hspace{0.5mm} \textup{dist} (\lambda,E)^{(\frac{p}{4} - 1 + \varepsilon)_{+}}}{\vert \lambda \vert^{\gamma}} \leq C_{0} \left( 1 + \frac{1}{\tau} \right)^{\gamma} K.
\end{equation}
\end{cor}
Theorem \ref{theo1} has immediate corollaries on sequences $(\lambda_{k}) \in \sigma_{d}(H)$ that converge to some $\lambda^{\ast} \in \sigma_{\textup{ess}} (H) = [\Lambda_{0},+\infty )$. Without loss of generality, we can consider a subsequence and assume that either (i) or (ii) below happens:
\begin{align*}
(\textup{i}) \quad \lambda^{\ast} & \in [\Lambda_{0},+\infty) \setminus E. \\
(\textup{ii}) \hspace{0.26cm} \lambda^{\ast} & \in E. 
\end{align*}
In case (i), the sequence $\big( \textup{dist} (\lambda_{k},E) \big)_{k}$ is positive and does not converge to 0. Then estimate \eqref{est0} implies that
\begin{equation}\label{est01}
\sum_{k} \vert \textup{Im} \hspace{0.5mm} \lambda_{k} \vert^{\frac{p}{2} + 1 + \varepsilon} < \infty.
\end{equation}
In case (ii), let us consider for example the $\lambda_{k}$ tending to a Landau level non-tangentially \big($i.e.$ $\vert \textup{Re} \hspace{0.5mm} \lambda_{k} - \lambda^{\ast} \vert \leq C \hspace{0.5mm} \vert \textup{Im} \hspace{0.5mm} \lambda_{k} \vert$ with some $C > 0$\big) and such that $\textup{dist} (\lambda_{k},E)$ is small enough. We then can claim that \eqref{est0} implies the estimate
\begin{equation}\label{est03}
\sum_{k} \textup{dist} (\lambda_{k},E)^{\frac{p}{2} + 1 + \varepsilon + (\frac{p}{4} - 1 + \varepsilon)_{+}} < \infty.
\end{equation}
So estimates \eqref{est01} and \eqref{est03} allows us to claim that $a\, priori$ the discrete eigenvalues of the operator $H$ are less densely distributed near Landau levels than elsewhere near it's essential spectrum.

Let us give some remarks about the three-dimensional case on selfadjoint perturbations $V$ satisfying the condition
\begin{equation}\label{est030}
V \leq 0, \quad C^{-1} \langle \textup{\textbf{x}} \rangle^{-\alpha} \leq \vert V(\textup{\textbf{x}}) \vert \leq C \hspace{0.5mm} \langle \textup{\textbf{x}} \rangle^{-\alpha}, \quad \alpha > 0
\end{equation}
with some constant $C > 1$. Let $(\lambda_{k})$ be a sequence of eigenvalues of $H$ accumulating to $\Lambda_{0}$ from the left. Suppose that it describes all eigenvalues $\lambda \in \sigma_{d}(H) \cap (\Lambda_{0}-r,\Lambda_{0})$ with some $r > 0$. Under some supplementary regularity assumptions on $V$ \big(see \cite[Theorem 1]{tam}\big), we have
\begin{equation}\label{est05}
\displaystyle \sum_{k} \textup{dist} \big(\lambda_{k},\Lambda_{0})^{p} = \int_{0}^{r} p\lambda^{p-1} N \big(\Lambda_{0}-\lambda, H\big) d\lambda < \infty
\end{equation}
for $p > 3/\alpha - 1/2$ if $\alpha < 2$ and $p(\alpha - 1) > 1$ if $\alpha > 2$. Here $N \big(\Lambda_{0}-\lambda, H\big)$ is the number of eigenvalues of $H$ less than $\Lambda_{0}-\lambda$ repeated according to their multiplicity. Hence in \eqref{est03}, conditions on $p$ are not optimal at least for selfadjoint perturbations $V$ of definite sign as above. Indeed, it can be checked that if the potential $V$ satisfies \eqref{est030} with $\alpha > 1/2$ \big(this is to compare to \eqref{eq0,52}\big), then $p/2 + 1 + \varepsilon > 3/\alpha - 1/2$ if $\alpha < 2$ and $p/2 + 1 + \varepsilon > 1/(\alpha - 1)$ if $\alpha > 2$.

\subsection{The $2d$-dimensional case}

The following theorem is proved in section 6 and concerns the $2d$-dimensional Schrödinger operator $H_{\perp}$ defined by \eqref{eq0.3}. We obtain a Lieb-Thirring type inequality for the discrete spectrum of the operator $H_{\perp}$.

\begin{theo}\label{theo3}
Let $H_{\perp} = H_{0,\perp} + V$ with $V$ satisfying \eqref{eq1,1} for $n = 2d$, $d \geq 1$. Assume that $F \in L^{p}\big(\mathbb{R}^{n}\big)$ with $p \geq 2 \big[ \frac{d}{2} \big] + 2$. Then the following holds
\begin{equation}\label{esta}
\sum_{\lambda \in \sigma_{d}(H_{\perp})} \frac{\textup{dist} \big(\lambda,E\big)^{p}}{\big( 1 + \vert \lambda \vert \big)^{2p}} \leq C_{1} \Vert F \Vert_{L^{p}}^{p} \big( 1 + \Vert V \Vert_{\infty} \big)^{2p},
\end{equation}
where $E$ is the set of Landau levels defined by \eqref{eq1,00}, and $C_{1} = C(p,b,d)$ is a constant depending on $p$, $b$ and $d$.
\end{theo}

Since for any $\tau > 0$ with $\vert \lambda \vert \geq \tau$ the lower bound \eqref{eq84} holds, we have the following corollary.

\begin{cor}
Under the assumptions and the notations of Theorem $\ref{theo3}$, the following bound holds for any $\tau > 0$
\begin{equation}\label{est85}
\displaystyle \sum_{\substack{\lambda \hspace{0.5mm} \in \hspace{0.5mm} \sigma_{d}(H_{\perp}) \\ \vert \lambda \vert \hspace{0.5mm} \geq \hspace{0.5mm} \tau}} \frac{\textup{dist} \big(\lambda,E\big)^{p}}{\vert \lambda \vert^{2p}} \leq C_{1} \left( 1 + \frac{1}{\tau} \right)^{2p} \Vert F \Vert_{L^{p}}^{p} \big( 1 + \Vert V \Vert_{\infty} \big)^{2p}.
\end{equation}
\end{cor}

Theorem \ref{theo3} has immediate corollaries on sequences $(\lambda_{k}) \in \sigma_{d}(H_{\perp})$ that converge to some $\lambda^{\ast} \in \sigma_{\textup{ess}} (H_{\perp}) = E$. Indeed, by \eqref{esta}
\begin{equation}\label{estb}
\sum_{k} \textup{dist} \big(\lambda_{k},E\big)^{p} < \infty,
\end{equation}
which $a\, priori$ means that the accumulation of the discrete eigenvalues near the Landau levels decreases with decreasing $p$.

Note that if the perturbation $V$ satisfies \eqref{eq0,5}, the finiteness of the sum in \eqref{estb} holds for $p > \max \hspace{0.4mm} (2d/m_{\perp},2)$, $m_{\perp} > 0$. However, it is convenient to mention that in the two-dimensional case $i.e.$ $d = 1$, if $V \geq 0$ is self-adjoint and satisfies \eqref{eq0,5} with $m_{\perp} > 1$, the condition $p > \max \hspace{0.4mm} (2/m_{\perp},1)$ is optimal for the finiteness of the sum in \eqref{estb}. This is a direct consequence of \cite[Theorem 2.6]{ra}, assuming some supplementary regularity assumptions on $V$. Indeed if $\lambda' \in (\Lambda_{j},\Lambda_{j+1})$, $j\geq0$ and $(\lambda_{k})$ an infinite sequence of discrete eigenvalues of $H_{\perp}$ accumulating to the Landau level $\Lambda_{j}$ from the right, then
\begin{equation}\label{estg}
\displaystyle \sum_{k} \textup{dist} \big(\lambda_{k},\Lambda_{j})^{p} = \int_{0}^{\lambda'-\Lambda_{j}} p\lambda^{p-1} N \big(\Lambda_{j}+\lambda,\lambda', H_{\perp}\big) d\lambda < \infty
\end{equation}
if and only if $pm_{\perp} > 2$. The quantity $N \big(\Lambda_{j}+\lambda,\lambda', H_{\perp}\big)$ is the number of eigenvalues of $H_{\perp}$ in $(\Lambda_{j}+\lambda,\lambda')$ repeated according to their multiplicity. Analogous result holds if we consider the eigenvalues of $H_{\perp} = H_{0,\perp} - V$ accumulating to $\Lambda_{j}$ from the left. Namely, \eqref{estg} remains valid if we replace $N \big(\Lambda_{j}+\lambda,\lambda', H_{\perp}\big)$ by $N \big(\lambda'',\Lambda_{j}-\lambda, H_{\perp}\big)$ with some $\lambda'' \in (\Lambda_{j-1},\Lambda_{j})$ if $j > 0$, or by $N \big(\Lambda_{j}-\lambda,H_{\perp}\big)$ if $j=0$ with $N \big(\Lambda_{j}-\lambda,H_{\perp}\big)$ defined as in \eqref{est05}.

Also in the two-dimensional case and self-adjoint perturbations $V \in C \big( \mathbb{R}^{2},\mathbb{R} \big)$ satisfying \eqref{eq0,51} with $m_{\perp} > 1$, we have the asymptotic property
\begin{equation}\label{este}
\displaystyle \sum_{k} \textup{dist} \big(\lambda_{k},\Lambda_{j})^{p} = o\Big(j^{-(p-1)/2}\Big), \quad j \longrightarrow \infty,
\end{equation}
for any $p \geq 1$ such that $p(m_{\perp} - 1) > 1$. This is a direct consequence of \cite[Lemma 1.5 and Theorem 1.6]{pus}.

\section{Estimate of the sandwiched resolvents of $H_{0}$ and $H$}

In order to estimate the resolvents of the operators $H_{0}$ and $H$, let us fix some notations. Denote by
\begin{equation}\label{eq1,31}
\mathbb{C}_{+} := \lbrace z \in \mathbb{C} : \textup{Im} \hspace{0.5mm} z > 0 \rbrace \quad \text{and} \quad \mathbb{C}_{-} := \lbrace z \in \mathbb{C} : \textup{Im} \hspace{0.5mm} z < 0 \rbrace.
\end{equation}
For $\Lambda_{j}$ defined by \eqref{eq1,00} and $\lambda \in \mathbb{C} \setminus [\Lambda_{0},+\infty)$, we have
\begin{equation}\label{eq1,3}
\displaystyle (H_{0} - \lambda)^{-1} = \sum_{j \in \mathbb{N}} p_{j} \otimes \Bigl( D_{x}^{2} + \Lambda_{j} - \lambda \Bigl)^{-1},
\end{equation}
where $p_{j}$ is the orthogonal projection onto $\ker \hspace{0.5mm} (H_{0,\perp} - \Lambda_{j})$. Recall that for $z \in \mathbb{C} \setminus [0,+\infty)$, the operator $\Bigl( D_{x}^{2} - z \Bigl)^{-1}$ admits the integral kernel 
\begin{equation}\label{eq1,4}
\frac{i \textup{e}^{i \sqrt{z} \vert x - x' \vert}}{2 \sqrt{z}},
\end{equation}
if the branch of $\sqrt{z}$ is chosen so that $\textup{Im} \hspace{0.5mm} \sqrt{z} > 0$. In the sequel, we assume that the perturbation $V$ satisfies assumption \eqref{eq1,1}. We have the following lemma.

\begin{lem}\label{lem1}
Let $n = 2d + 1$, $d \geq 1$ and $\lambda \in \mathbb{C} \setminus [\Lambda_{0},+\infty)$. Assume that $F \in L^{p}\big(\mathbb{R}^{n}\big)$ with $p \geq 2 \big[ \frac{d}{2} \big] + 2$ and $G \in \big( L^{2} \cap L^{\infty} \big) \big{(} \mathbb{R} \big{)}$. Then there exists a constant $C = C(p,b,d)$ such that 
\begin{equation}\label{est4}
\left\Vert F (H_{0} - \lambda )^{-1} G \right\Vert_{p}^{p} \leq \frac{C (1 + \vert \lambda \vert)^{d + \frac{1}{2}} K_{1}}{\textup{dist} \big(\lambda,[\Lambda_{0},+\infty)\big)^{\frac{p}{2}} \hspace{0.5mm} \textup{dist} (\lambda,E)^{\frac{p}{4}}},
\end{equation}
where $E$ is the set of Landau levels defined by \eqref{eq1,00} and
\begin{equation}\label{eq1,50}
K_{1} := \Vert F \Vert_{L^{p}}^{p} \big{(} \Vert G \Vert_{L^{2}} + \Vert G \Vert_{L^{\infty}} \big{)}^{p}.
\end{equation} 
\end{lem}

\hspace{-0.54cm} \begin{prof}
It suffices to prove the case $\lambda \in \mathbb{C}_{+}$. Constants are generic (changing from a relation to another).
\\\\
$(\textup{i})$: We show that \eqref{est4} holds if $p$ is even. Let us first prove that if $\textup{Re} \hspace{0.5mm} \lambda < \Lambda_{0}$, then
\begin{equation}\label{est2}
\left\Vert F (H_{0} - \lambda )^{-1} G \right\Vert_{p}^{p} \leq \frac{C (1 + \vert \lambda \vert)^{d + \frac{1}{2}} K_{1}} {\textup{dist} (\lambda,E)^{\frac{3p}{4}}}.
\end{equation} 
Using the identity
$$(H_{0} - \lambda)^{-1} - \bigl( H_{0} + 1 + \vert \lambda \vert \bigr)^{-1} = \bigl( H_{0} + 1 + \vert \lambda \vert \bigr)^{-1} (1 + \vert \lambda \vert +\lambda) (H_{0} - \lambda)^{-1},$$
we get $$(H_{0} - \lambda)^{-1} = \bigl( H_{0} + 1 + \vert \lambda \vert \bigr)^{-1} \bigl( (1 + \vert \lambda \vert + \lambda) (H_{0} - \lambda)^{-1} + I \bigr).$$ 
Then 
\begin{equation}\label{eq3.1}
\begin{split}
\left\Vert F (H_{0} - \lambda)^{-1} G \right\Vert^{p}_{p} \leq & \left\Vert F \bigl( H_{0} + 1 + \vert \lambda \vert \bigr)^{-1} \right\Vert^{p}_{p} \\
& \times \left\Vert \Bigl( (1 + \vert \lambda \vert + \lambda) (H_{0} - \lambda)^{-1} + I \Bigr) G \right\Vert^{p}
\end{split}
\end{equation}
Since $p$ is even, we can apply the diamagnetic inequality for the $S_{p}$ class operators \big(\cite[Theorem 2.3]{avr} and \cite[Theorem 2.13]{sim}\big). We get
$$\left\Vert F \bigl( H_{0} + 1 + \vert \lambda \vert \bigr)^{-1} \right\Vert^{p}_{p} \leq \left\Vert F (-\Delta + 1 + \vert \lambda \vert)^{-1} \right\Vert^{p}_{p}.$$
By \cite[Theorem 4.1]{sim},
$$\left\Vert F (-\Delta + 1 + \vert \lambda \vert)^{-1} \right\Vert^{p}_{p} \leq C(p) \left\Vert F \right\Vert_{L^{p}}^{p} \left\Vert \Bigl( \vert \cdot \vert^{2} + 1 + \vert \lambda \vert \Bigr)^{-1} \right\Vert_{L^{p}}^{p}.$$
Since 
$$\left\Vert \Bigl( \vert \cdot \vert^{2} + 1 + \vert \lambda \vert \Bigr)^{-1} \right\Vert_{L^{p}}^{p} = \displaystyle C \int_{0}^{\infty} \frac{r^{2d}dr}{({r}^{2} + 1 + \vert \lambda \vert)^{p}} = C(p) \frac{(1 + \vert \lambda \vert)^{d + \frac{1}{2}}}{(1 + \vert \lambda \vert)^{p}},$$
then 
\begin{equation}\label{eq3.2}
\left\Vert F \bigl( H_{0} + 1 + \vert \lambda \vert \bigr)^{-1} \right\Vert^{p}_{p} \leq C(p) \frac{(1 + \vert \lambda \vert)^{d + \frac{1}{2}}}{(1 + \vert \lambda \vert)^{p}} \left\Vert F \right\Vert_{L^{p}}^{p}.
\end{equation}
Otherwise,
\begin{equation}\label{eq3.3}
\begin{aligned}
& \left\Vert \Bigl( (1 + \vert \lambda \vert + \lambda) (H_{0} - \lambda)^{-1} + I \Bigr) G \right\Vert^{p} \\
& \leq \Big{(} C \bigl( 1 + \vert \lambda \vert \bigr) \left\Vert (H_{0} - \lambda)^{-1} G \right\Vert + \left\Vert G \right\Vert_{L^{\infty}} \Big{)}^{p}.
\end{aligned}
\end{equation}
By \eqref{eq1,3} we have
\begin{equation}\label{eq3.31b}
\begin{split}
\left\Vert (H_{0} - \lambda)^{-1} G \right\Vert & = \displaystyle \underbrace{\left\Vert \sum_{j \in \mathbb{N}} p_{j} \otimes \Bigl( D_{x}^{2} + \Lambda_{j} - \lambda \Bigl)^{-1} G \right\Vert}_{\textup{direct sum}}\\
& \leq C \hspace{0.5mm} \sup_{j} \left\Vert \Bigl( D_{x}^{2} + \Lambda_{j} - \lambda \Bigl)^{-1} G \right\Vert,
\end{split}
\end{equation}
and moreover
\begin{equation}\label{eq3.4}
\begin{split}
\left\Vert \Bigl( D_{x}^{2} + \Lambda_{j} - \lambda \Bigl)^{-1} G \right\Vert & \leq \left\Vert \Bigl( D_{x}^{2} + \Lambda_{j} - \lambda \Bigl)^{-1} G \right\Vert_{2}\\
& = \left\Vert G \Bigl( D_{x}^{2} + \Lambda_{j} - \overline{\lambda} \Bigl)^{-1} \right\Vert_{2}.
\end{split}
\end{equation}
As above, by \cite[Theorem 4.1]{sim}, 
$$\left\Vert G \Bigl( D_{x}^{2} + \Lambda_{j} - \overline{\lambda} \Bigl)^{-1} \right\Vert_{2}^{2} \leq C \hspace{0.5mm} \left\Vert G \right\Vert_{L^{2}}^{2} \left\Vert \Bigl( \vert \cdot \vert^{2} + \Lambda_{j} - \overline{\lambda} \Bigl)^{-1} \right\Vert_{L^{2}(\mathbb{R})}^{2}.$$
Since $\textup{Re} \hspace{0.5mm} \lambda < \Lambda_{0}$, then for any $r \in \mathbb{R}$ we have 
$$\left\vert r^{2} + \Lambda_{j} - \overline{\lambda} \right\vert^{2} \geq r^{4} + \left\vert \overline{\lambda} - \Lambda_{j} \right\vert^{2}.$$
This implies that
$$\left\Vert \Bigl( \vert \cdot \vert^{2} + \Lambda_{j} - \overline{\lambda} \Bigl)^{-1} \right\Vert_{L^{2}}^{2} \leq \left\Vert \Bigl( \vert \cdot \vert^{4} + \vert \overline{\lambda} - \Lambda_{j} \vert^{2} \Bigl)^{-1}  \right\Vert_{L^{1}(\mathbb{R})} \leq \frac{C}{\vert \overline{\lambda} - \Lambda_{j} \vert^{\frac{3}{2}}},$$
so that finally
\begin{equation}\label{eq3.5}
\left\Vert (H_{0} - \lambda)^{-1} G \right\Vert \leq \frac{C \left\Vert G \right\Vert_{L^{2}}}{\textup{dist} (\lambda,E)^{\frac{3}{4}}}.
\end{equation}
By \eqref{eq3.3} and \eqref{eq3.5}, we have
\begin{equation}\label{eq3.6}
\begin{split}
\left\Vert \Bigl( (1 + \vert \lambda \vert + \lambda) (H_{0} - \lambda)^{-1} + I \Bigr) G \right\Vert^{p} & \leq \left( \frac{C \left\Vert G \right\Vert_{L^{2}} \bigl( 1 + \vert \lambda \vert \bigr)}{\textup{dist} (\lambda,E)^{\frac{3}{4}}} + \left\Vert G \right\Vert_{L^{\infty}} \right)^{p}\\
& \leq \frac{C(p,b,d) \Bigl( \left\Vert G \right\Vert_{L^{2}} + \left\Vert G \right\Vert_{L^{\infty}} \Bigr)^{p} \bigl( 1 + \vert \lambda \vert \bigr)^{p}} {\textup{dist} (\lambda,E)^{\frac{3p}{4}}},
\end{split}
\end{equation}
and \eqref{est2} follows from \eqref{eq3.1}, \eqref{eq3.2} and \eqref{eq3.6}.
\\

Now let us prove that if $\textup{Re} \hspace{0.5mm} \lambda \geq \Lambda_{0}$, then
\begin{equation}\label{est3}
\left\Vert F (H_{0} - \lambda )^{-1} G \right\Vert_{p}^{p} \leq \frac{C (1 + \vert \lambda \vert)^{d + \frac{1}{2}} K_{1}}{\vert \textup{Im} \hspace{0.5mm} \lambda \vert^{\frac{p}{2}} \hspace{0.5mm} \textup{dist} (\lambda,E)^{\frac{p}{4}}}.
\end{equation}
From \eqref{eq3.4}, we compute the Hilbert-Schmidt norm of $G \Bigl( D_{x}^{2} + \Lambda_{j} - \overline{\lambda} \Bigl)^{-1}$ with the help of its integral kernel \eqref{eq1,4}. Thus for $\textup{Im} \hspace{0.5mm} \sqrt{\overline{\lambda} - \Lambda_{j}} > 0$, we have
$$\left\Vert G \Bigl( D_{x}^{2} + \Lambda_{j} - \overline{\lambda} \Bigl)^{-1} \right\Vert_{2}^{2} = \frac{C \Vert G \Vert^{2}_{L^{2}}}{\textup{Im} \hspace{0.3mm} \sqrt{\overline{\lambda} - \Lambda_{j}} \hspace{0.3mm} \vert \overline{\lambda} - \Lambda_{j} \vert}.$$
Now from $$\textup{Im} \hspace{0.3mm} \overline{\lambda} = \textup{Im} \hspace{0.3mm} (\overline{\lambda} - \Lambda_{j}) = 2 \hspace{0.3mm} \textup{Im} \hspace{0.3mm} \sqrt{\overline{\lambda} - \Lambda_{j}} \hspace{0.3mm} \textup{Re} \hspace{0.3mm} \sqrt{\overline{\lambda} - \Lambda_{j}},$$ 
we deduce that
\begin{equation}\label{eq3.7}
\begin{split}
\left\Vert G \Bigl( D_{x}^{2} + \Lambda_{j} - \overline{\lambda} \Bigl)^{-1} \right\Vert_{2}^{2} & = \frac{C \Vert G \Vert^{2}_{L^{2}} \hspace{0.3mm} \textup{Re} \hspace{0.3mm} \sqrt{\overline{\lambda} - \Lambda_{j}}}{\vert \textup{Im} \hspace{0.3mm} \overline{\lambda} \vert \hspace{0.2mm} \vert \overline{\lambda} - \Lambda_{j} \vert} \\
& \leq \frac{C \Vert G \Vert^{2}_{L^{2}}}{\vert \textup{Im} \hspace{0.3mm} \lambda \vert \hspace{0.3mm} \textup{dist} (\lambda,E)^{\frac{1}{2}}}.
\end{split}
\end{equation}
Combining \eqref{eq3.31b}, \eqref{eq3.4} and \eqref{eq3.7}, we get
\begin{equation}\label{eq3.8}
\left\Vert (H_{0} - \lambda)^{-1} G \right\Vert \leq \frac{C \Vert G \Vert_{L^{2}}}{\vert \textup{Im} \hspace{0.3mm} \lambda \vert^{\frac{1}{2}} \hspace{0.3mm} \textup{dist} (\lambda,E)^{\frac{1}{4}}}.
\end{equation}
Finally by \eqref{eq3.3} and \eqref{eq3.8},
\begin{equation}\label{eq3.9}
\begin{split}
\left\Vert \Bigl( (1 + \vert \lambda \vert + \lambda) (H_{0} - \lambda)^{-1} + I \Bigr) G \right\Vert^{p} & \leq \frac{C(p) \Bigl( \left\Vert G \right\Vert_{L^{2}} + \left\Vert G \right\Vert_{L^{\infty}} \Bigr)^{p} \bigl( 1 + \vert \lambda \vert \bigr)^{p}}{\vert \textup{Im} \hspace{0.3mm} \lambda \vert^{\frac{p}{2}} \hspace{0.3mm} \textup{dist} (\lambda,E)^{\frac{p}{4}}}.
\end{split}
\end{equation}
Now \eqref{est3} follows from \eqref{eq3.1}, \eqref{eq3.2} and \eqref{eq3.9}. Hence estimates \eqref{est2} and \eqref{est3} show that \eqref{est4} holds if $p$ is even.
\\\\
$(\textup{ii})$: We prove that \eqref{est3} holds for any $p \geq 2 \big[ \frac{d}{2} \big] + 2$ using interpolation method. Clearly if $p$ is as above, there exists even integers $p_{0} < p_{1}$ such that $p \in (p_{0},p_{1})$ with  $p_{0} > d + 1/2$. Now choose $s \in (0,1)$ such that $\frac{1}{p} = \frac{1-s}{p_{0}} + \frac{s}{p_{1}}$. For $i = 0$, $1$ consider the operator
$$L^{p_{i}}\big(\mathbb{R}^{n}\big) \ni F \overset{T}{\longmapsto} F (H_{0} - \lambda )^{-1} G \in S_{p_{i}}.$$
By (i) proved above, estimate \eqref{est4} holds for any $F \in L^{p_{i}}\big(\mathbb{R}^{n}\big)$. Let $C_{i} = C(p_{i},b,d)$ be the constant in \eqref{est4} and define $$C(\lambda,p_{i},d) := \frac{C_{i}^{\frac{1}{p_{i}}} \left( (1 + \vert \lambda \vert)^{d + \frac{1}{2}} \right)^{\frac{1}{p_{i}}} \Big{(} \Vert G \Vert_{L^{2}} + \Vert G \Vert_{L^{\infty}} \Big{)}}{\textup{dist} \big(\lambda,[\Lambda_{0},+\infty)\big)^{\frac{1}{2}} \hspace{0.5mm} \textup{dist} (\lambda,E)^{\frac{1}{4}}}.$$
Then by \eqref{est4} for $(\textup{i})$ proved above, we have $\Vert T \Vert \leq C(\lambda,p_{i},d)$ for any $i = 0$, $1$. Using the Riesz-Thorin Theorem \big(see e.g. \cite[sub. 5 of chap. 6]{fol}, \cite{rie}, \cite{tho}, \cite[chap. 2]{lun}\big), we can interpolate between $p_{0}$ and $p_{1}$ to get the extension $T : L^{p}\big(\mathbb{R}^{2d + 1}\big) \longrightarrow S_{p}$ with
\begin{align*}
\Vert T \Vert & \leq C(\lambda,p_{0},d)^{1-s} C(\lambda,p_{1},d)^{s} \\
& \leq \frac{C(p,b,d) \left( (1 + \vert \lambda \vert)^{d + \frac{1}{2}} \right)^{\frac{1}{p}} \Big{(} \Vert G \Vert_{L^{2}} + \Vert G \Vert_{L^{\infty}} \Big{)}}{\textup{dist} \big(\lambda,[\Lambda_{0},+\infty)\big)^{\frac{1}{2}} \hspace{0.5mm} \textup{dist} (\lambda,E)^{\frac{1}{4}}}.
\end{align*}
In particular for any $F \in L^{p}\big(\mathbb{R}^{n}\big)$, we have
\begin{equation}\label{eq3.92}
\Vert T (F) \Vert_{p} \leq \frac{C(p,b,d) \left( (1 + \vert \lambda \vert)^{d + \frac{1}{2}} \right)^{\frac{1}{p}} \Big{(} \Vert G \Vert_{L^{2}} + \Vert G \Vert_{L^{\infty}} \Big{)}}{\textup{dist} \big(\lambda,[\Lambda_{0},+\infty)\big)^{\frac{1}{2}} \hspace{0.5mm} \textup{dist} (\lambda,E)^{\frac{1}{4}}} \left\Vert F \right\Vert_{L^{p}},
\end{equation}
which implies estimate \eqref{est4}, and the lemma is proved.
\end{prof}

Now let $\lambda_{0}$ be such that
\begin{equation}\label{eq3.91}
\min \Big( \vert \textup{Im} \hspace{0.3mm} \lambda_{0} \vert,\textup{dist} \big(\lambda_{0},\overline{N(H)}\big) \Big) \geq 1 + \Vert V \Vert_{\infty}.
\end{equation}
We have the following lemma.

\begin{lem}\label{lem11}
Assume that $\lambda_{0}$ satisfies condition \eqref{eq3.91}.  Under the assumptions of Lemma $\ref{lem1}$, there exists a constant $C = C(p)$ such that
\begin{equation}\label{est41}
\left\Vert F (H - \lambda_{0})^{-1} G \right\Vert_{p}^{p} \leq C (1 + \vert \lambda_{0} \vert)^{d + \frac{1}{2}} K_{2},
\end{equation}
where the constant $K_{2}$ is defined by
\begin{equation}\label{eq1,500}
K_{2} := \Vert F \Vert_{L^{p}}^{p} \Vert G \Vert_{L^{\infty}}^{p}.
\end{equation} 
\end{lem}

\hspace{-0.54cm} \begin{prof}
Constants are generic (changing from a relation to another). From
$$(H - \lambda_{0})^{-1} = (H_{0} - \lambda_{0})^{-1} (H_{0} - \lambda_{0}) (H - \lambda_{0})^{-1},$$
we deduce that 
\begin{equation}\label{est42}
\begin{split}
\left\Vert F (H - \lambda_{0})^{-1} G \right\Vert_{p}^{p} & \leq \left\Vert F (H - \lambda_{0})^{-1} \right\Vert_{p}^{p} \Vert G \Vert_{L^{\infty}}^{p} \\
& \leq \left\Vert F (H_{0} - \lambda_{0})^{-1} \right\Vert_{p}^{p} \left\Vert (H_{0} - \lambda_{0}) (H - \lambda_{0})^{-1} \right\Vert^{p} \Vert G \Vert_{L^{\infty}}^{p}.
\end{split}
\end{equation}
Using similar method to that of the proof of Lemma \ref{lem1}, we can show that
\begin{equation}\label{est43}
\left\Vert F (H_{0} - \lambda_{0})^{-1} \right\Vert_{p}^{p} \leq \frac{C (1 + \vert \lambda_{0} \vert)^{d + \frac{1}{2}} \Vert F \Vert_{L^{p}}^{p}}{\vert \textup{Im} \hspace{0.3mm} \lambda_{0} \vert^{p}}.
\end{equation}
So \eqref{est43} together with condition \eqref{eq3.91} on $\lambda_{0}$ give finally
\begin{equation}\label{est45}
\left\Vert F (H_{0} - \lambda_{0})^{-1} \right\Vert_{p}^{p} \leq C (1 + \vert \lambda_{0} \vert)^{d + \frac{1}{2}} \Vert F \Vert_{L^{p}}^{p}.
\end{equation}
Otherwise, we have
$$\left\Vert (H_{0} - \lambda_{0}) (H - \lambda_{0})^{-1} \right\Vert = \left\Vert I - V(H - \lambda_{0})^{-1} \right\Vert \leq 1 + \Vert V \Vert_{\infty} \left\Vert (H - \lambda_{0})^{-1} \right\Vert.$$
By \cite[Lemma 9.3.14]{dav}, 
$$\left\Vert (H - \lambda_{0})^{-1} \right\Vert \leq \frac{1}{\textup{dist} \big(\lambda_{0},\overline{N(H)}\big)}.$$
The assumption on $\lambda_{0}$ implies then that
\begin{equation}\label{est44}
\left\Vert (H_{0} - \lambda_{0}) (H - \lambda_{0})^{-1} \right\Vert \leq 2.
\end{equation}
Now \eqref{est41} follows from \eqref{est42}, \eqref{est45} and \eqref{est44}.
\end{prof}

\section{Preliminaries}

\subsection{About the holomorphic function $f(\lambda)$}

In this subsection, we construct as in \cite{dem} a holomorphic function $f : \mathbb{C} \setminus [\Lambda_{0},+\infty) \rightarrow \mathbb{C}$ whose zeros are the $\lambda \in \sigma_{d}(H)$. 

Let $\lambda \in \mathbb{C} \setminus [\Lambda_{0},+\infty)$. We have the identity
\begin{equation}\label{eq3.10}
(H - \lambda)(H_{0} - \lambda)^{-1} = I + V(H_{0} - \lambda)^{-1}.
\end{equation}
LHS of \eqref{eq3.10} is not invertible if and only if $H - \lambda$ is not invertible. Since $V$ is a relatively compact perturbation, this happens if and only if $\lambda \in \sigma_{d}(H)$. So defining 
\begin{equation}\label{eq3.11}
T(\lambda) = V(H_{0} - \lambda)^{-1},
\end{equation}
we get for $\lambda \in \mathbb{C} \setminus [\Lambda_{0},+\infty)$ that
\begin{equation}\label{eq3.12}
\lambda \in \sigma_{d}(H) \Leftrightarrow I + T(\lambda) \hspace{0.2cm} \textup{is not invertible}.
\end{equation}
Otherwhise, assumption \eqref{eq1,1} for $n = 2d + 1$ on the potential $V$ implies that exists a bounded operator $\mathcal{V}$ such that for any $\textup{\textbf{x}} = (X_{\perp},x) \in \mathbb{R}^{n}$, $V(\textup{\textbf{x}}) = \mathcal{V} F(\textup{\textbf{x}}) G(x)$. Thus as in proof of Lemma \ref{lem1}, we can show that $T(\lambda) \in S_{p}$ for any $p \geq 2$. Let $\textup{det}_{\lceil p \rceil} \big{(} I + T(\lambda) \big{)}$ be the regularized determinant defined by
\begin{equation}\label{eq3.121}
\small{\textup{det}_{\lceil p \rceil} \big( I + T(\lambda) \big) := \prod_{\mu \in \sigma \big( T(\lambda) \big)} \left[ (1 + \mu) \exp \left( \sum_{k=1}^{\lceil p \rceil-1} \frac{(-\mu)^{k}}{k} \right) \right]},
\end{equation}
where $\lceil p \rceil := \min \lbrace n \in \mathbb{N} : n \geq p \rbrace$. Hence \eqref{eq3.12} can be formulated as \big(see e.g. \cite[chap. 9]{sim}\big)
\begin{equation}\label{eq3.13}
\lambda \in \sigma_{d}(H) \Leftrightarrow \textup{det}_{\lceil p \rceil} \big{(} I + T(\lambda) \big{)} = 0.
\end{equation}
Let us define
\begin{equation}\label{eq3.131}
f(\lambda) := \textup{det}_{\lceil p \rceil} \big{(} I + T(\lambda) \big{)}.
\end{equation} 
The function $f$ is holomorphic on $\mathbb{C} \setminus [\Lambda_{0},+\infty)$ and then 
\begin{equation}\label{eq3.14}
\sigma_{d}(H) = \lbrace \lambda \in \mathbb{C} \setminus [\Lambda_{0},+\infty) : f(\lambda) = 0 \rbrace.
\end{equation}
Moreover, the algebraic multiplicity of $\lambda \in \sigma_{d}(H)$ is equal to the order of $\lambda$ as zero of the function $f$.
\\\\
Let us proof the following lemma on $f(\lambda)$.

\begin{lem}\label{lem2}
Let $\lambda \in \mathbb{C} \setminus [\Lambda_{0},+\infty)$ and suppose that $\lambda_{0}$ satisfies \eqref{eq3.91}. Under the assumptions of Lemma $\ref{lem1}$, there exists $C = C(p,b,d)$ such that
\begin{equation}\label{eq3.15}
\small{\log \left\vert \frac{f(\lambda)}{f(\lambda_{0})} \right\vert \leq \frac{\Gamma_{p} C (1 + \vert \lambda \vert)^{d+\frac{1}{2}} K_{1}}{\textup{dist} \big(\lambda,[\Lambda_{0},+\infty)\big)^{\frac{p}{2}} \hspace{0.5mm} \textup{dist} (\lambda,E)^{\frac{p}{4}}} + \Gamma_{p} C (1 + \vert \lambda_{0} \vert)^{d + \frac{1}{2}} K_{1}},
\end{equation}
where $\Gamma_{p}$ is some positive constant and $K_{1}$ is defined by \eqref{eq1,50}.
\end{lem}

\hspace{-0.54cm} \begin{prof}
First, write $\frac{f(\lambda)}{f(\lambda_{0})} = f(\lambda) \cdot f(\lambda_{0})^{-1}$. Since
$$f(\lambda_{0})^{-1} = \textup{det}_{\lceil p \rceil} \Big{(} I + V(H_{0} - \lambda_{0})^{-1} \Big{)}^{-1},$$
then passing to the inverse in \eqref{eq3.10}, we get
$$f(\lambda_{0})^{-1} = \textup{det}_{\lceil p \rceil} \Big{(} I - V(H - \lambda_{0})^{-1} \Big{)}.$$
This implies that
\begin{equation}\label{eq3.151}
\begin{split}
\small{\left\vert \frac{f(\lambda)}{f(\lambda_{0})} \right\vert} & = \small{\left\vert \textup{det}_{\lceil p \rceil} \Big{(} I + V(H_{0} - \lambda)^{-1} \Big{)} \right\vert \cdot \left\vert \textup{det}_{\lceil p \rceil} \Big{(} I - V(H - \lambda_{0})^{-1} \Big{)} \right\vert} \\
& = \small{\left\vert \textup{det}_{\lceil p \rceil} \Big{(} I + \mathcal{V} F (H_{0} - \lambda)^{-1} G \Big{)} \right\vert \cdot \left\vert \textup{det}_{\lceil p \rceil} \Big{(} I - \mathcal{V} F (H - \lambda_{0})^{-1} G \Big{)} \right\vert}.
\end{split}
\end{equation}
Otherwise, for any $A \in S_{\lceil p \rceil}$, we have the estimate \big(see e.g. \cite{sim}\big)
$$\left\vert \textup{det}_{\lceil p \rceil} (I + A) \right\vert \leq \textup{e}^{\Gamma_{p} \Vert A \Vert_{p}^{p}}.$$
Thus using \eqref{eq3.151}, we get
\begin{equation}\label{eq3.152}
\left\vert \frac{f(\lambda)}{f(\lambda_{0})} \right\vert \leq \textup{e}^{C \hspace{0.02cm} \Gamma_{p} \left( \Vert F (H_{0} - \lambda)^{-1} G \Vert_{p}^{p} + \Vert F (H - \lambda_{0})^{-1} G \Vert_{p}^{p} \right)}.
\end{equation}
So Lemma \ref{lem1} and Lemma \ref{lem11} together with the inequality $K_{2} < K _{1}$ give \eqref{eq3.15}.
\end{prof}

In what follows below, we use a theorem about zeros of holomorphic functions in the unit disk $\mathbb{D} := \lbrace \vert z \vert < 1 \rbrace$ to study the zeros of the function $f$ in $\mathbb{C} \setminus [\Lambda_{0},+\infty)$. Let us recall that in our case the Landau levels $\Lambda_{j}$ defined by \eqref{eq1,00} play the role of thresholds in $[\Lambda_{0},+\infty)$. So to study the zeros of the function $f$ in $\mathbb{C} \setminus [\Lambda_{0},+\infty)$ or equivalently the discrete eigenvalues of the operator $H$ in $\mathbb{C} \setminus [\Lambda_{0},+\infty)$, we use a local approach by transforming locally the problem to $\mathbb{D}$ using a conformal map.

\subsection{About the conformal map $\varphi(z)$}

Let $\Pi = \mathcal{R}(\lambda_{1},\lambda_{2},\lambda_{3},\lambda_{4})$ be a rectangle (or a square) with vertices $\lambda_{1}$, $\lambda_{4} \in \mathbb{R}$, and $\lambda_{2}$, $\lambda_{3} \in \mathbb{C}_{+}$ or $\lambda_{2}$, $\lambda_{3} \in \mathbb{C}_{-}$ (see the figure below). It is well known \big(see e.g. \cite[Theorem 1, p. 176]{lav}\big), that there exists a conformal map $\varphi : \mathbb{D} \rightarrow \Pi$ given by Schwarz-Christoffel formula. Denote by 
$$\mathbb{T} := \partial \mathbb{D} = \lbrace z \in \mathbb{C} : \vert z \vert = 1 \rbrace,$$
and let $z_{j} \in \mathbb{T}$ be the points such that $\varphi(z_{j}) = \lambda_{j}$, for $1 \leq j \leq 4$. Since $\Pi$ is a rectangle, the map $\varphi$ satisfies
\begin{equation}\label{eq3.16}
\varphi'(z) = \frac{C}{(z - z_{1})^{\frac{1}{2}}(z - z_{2})^{\frac{1}{2}}(z - z_{3})^{\frac{1}{2}}(z - z_{4})^{\frac{1}{2}}}
\end{equation}
where $C$ is a constant.

\begin{center}
\tikzstyle{ddEncadre}=[densely dotted]
\begin{tikzpicture}[scale = 0.8]
\draw (0,0) circle(1.5);
\node at (0,0) {\tiny{$\times$}};
\node at (1.2,0.9) {\tiny{$\bullet$}};
\node at (1.5,1.1) {$z_{4}$};
\node at (1.2,-0.9) {\tiny{$\bullet$}};
\node at (1.5,-1.1) {$z_{1}$};
\node at (-1.2,-0.9) {\tiny{$\bullet$}};
\node at (-1.5,-1.1) {$z_{2}$};
\node at (-1.2,0.9) {\tiny{$\bullet$}};
\node at (-1.5,1.1) {$z_{3}$};
\node at (0,1.8) {$\mathcal{L}_{2,z}$};
\node at (0,-1.8) {$\mathcal{L}_{4,z}$};
\node at (1.94,0) {$\mathcal{L}_{1,z}$};
\node at (-1.9,0) {$\mathcal{L}_{3,z}$};

\draw (4,-1.5) -- (10,-1.5);
\node at (10.2,-1.5) {$\mathbb{R}$};
\draw (5,-1.5) -- (5,1.2) -- (9,1.2) -- (9,-1.5);
\node at (9.1,-1.8) {$\lambda_{4}$};
\node at (5,-1.8) {$\lambda_{1}$};
\node at (9.2,1.5) {$\lambda_{3}$};
\node at (5,1.5) {$\lambda_{2}$};
\node at (7,-1.8) {$\mathcal{L}_{1,\lambda}$};
\node at (7,1.5) {$\mathcal{L}_{3,\lambda}$};
\node at (9.45,0) {$\mathcal{L}_{2,\lambda}$};
\node at (4.6,0) {$\mathcal{L}_{4,\lambda}$};

\node at (-0.6,0.5) {$z$};
\node at (6.5,-0.7) {$\lambda$};

\draw [ddEncadre] [->,>=latex] (-0.4,0.5) -- (6.5,0.5);
\node at (3.2,0.7) {$\varphi : \mathbb{D} \rightarrow \Pi$};
\draw [ddEncadre] [->,>=latex] (6.3,-0.7) -- (-0.6,-0.7);
\node at (3.2,-0.5) {$\varphi^{-1}$};
\end{tikzpicture}

\vspace*{0.2cm}

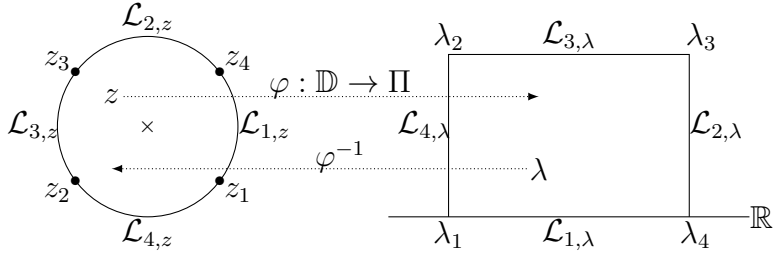
\textsc{Figure $1$.} Transformation of the unit disk by the conformal map $\varphi(z) = \lambda$.
\end{center}

We write $\lambda = \varphi(z) = \lambda(z)$ or $z = \varphi^{-1}(\lambda) = z(\lambda)$. We have $\varphi(\mathcal{L}_{j,z}) = \mathcal{L}_{j,\lambda}$ and $\partial \Pi = \bigcup_{j} \mathcal{L}_{j,\lambda}$. We set also $\mathcal{F}_{z} :=  \lbrace z_{j} \rbrace_{j}$ and $\mathcal{F}_{\lambda} :=  \lbrace \lambda_{j} \rbrace_{j}$ so that $\varphi(\mathcal{F}_{z}) = \mathcal{F}_{\lambda}$. Elsewhere in this paper, $\mathscr{P} \simeq \mathscr{Q}$ means that there exist constants $C_{1}$, $C_{2}$ such that 
\begin{equation}\label{eq3.161}
0 < C_{1} \leq \frac{\mathscr{P}}{\mathscr{Q}} \leq C_{2} < \infty.
\end{equation}

\begin{lem}\label{lem3}
For $z \in \mathbb{D}$ and $\lambda \in \Pi$, the following holds,
$$\small{\textup{dist} (\lambda,\partial\Pi) \simeq \textup{dist} (z,\mathbb{T}) \frac{1}{\textup{dist} (z,\mathcal{F}_{z})^{\frac{1}{2}}}}$$ or equivalently $$\small{\textup{dist} (z,\mathbb{T}) \simeq \textup{dist} (\lambda,\partial\Pi) \hspace{0.6mm} \textup{dist} (\lambda,\mathcal{F}_{\lambda})}.$$
\end{lem}

\hspace{-0.54cm} \begin{prof}
Follows directly from \cite[Corollary 1.4]{pom} and \eqref{eq3.16}.
\end{prof}

In the sequel, we are interested in the same quantities where $\textup{dist} (z,\mathbb{T})$ and $\textup{dist} (\lambda,\partial\Pi)$ are replaced respectively by $\textup{dist} (z,\mathcal{L}_{1,z})$ and $\textup{dist} (\lambda,\mathcal{L}_{1,\lambda})$. As in Lemma \ref{lem3}, we have
\begin{equation}\label{eq3.17}
\begin{aligned}
\small{\textup{dist} (\lambda,\mathcal{L}_{1,\lambda})} & \small{\simeq \textup{dist} (z,\mathcal{L}_{1,z}) \frac{1}{\textup{dist} \big( z,\lbrace z_{1},z_{4} \rbrace \big)^{\frac{1}{2}}}} \\
\small{\textup{dist} (z,\mathcal{L}_{1,z})} & \small{\simeq \textup{dist} (\lambda,\mathcal{L}_{1,\lambda}) \hspace{0.6mm} \textup{dist} \big( \lambda,\lbrace \lambda_{1},\lambda_{4} \rbrace \big)}.
\end{aligned}
\end{equation}

\subsection{About a theorem in \cite{bor}}

The following result by A. Borichev, L. Golinskii and S. Kupin is proved in \cite{bor}. It gives an estimate on zeros of holomorphic functions in the unit disk $\mathbb{D}$.

\begin{theo}\label{theo2}
Let $h$ be a holomorphic function in the unit disk $\mathbb{D}$ with $h(0) = 1$. Assume that $h$ satisfies a bound of the form 
$$\log \vert h(z) \vert \leq K_{0} \frac{1}{(1 - \vert z \vert)^{\alpha}} \prod_{j=1}^{N} \frac{1}{\vert z - \xi_{j} \vert^{\beta_{j}}},$$
where $\xi_{j} \in \mathbb{T}$ and $\alpha$, $\beta_{j} \geq 0$. Let $\tau > 0$. Then the zeros of $h$ satisfy the inequality
$$\sum_{\left\lbrace h(z)=0 \right\rbrace} (1 - \vert z \vert)^{\alpha + 1 + \tau} \prod_{j=1}^{N} \vert z - \xi_{j} \vert^{(\beta_{j} - 1 + \tau)_{+}} \leq C \big{(} \alpha,\lbrace \beta_{j} \rbrace,\lbrace \xi_{j} \rbrace,\tau \big{)} K_{0}.$$
\end{theo}

\section{Bounds on the discrete eigenvalues of $H$}

\subsection{Local bound}

Let us recall that the $\Lambda_{j}$ are the Landau levels given by \eqref{eq1,00}. In this subsection, we prove a bound on $\lambda \in \sigma_{d} (H)$ in a rectangle $\Pi_{j}$ containing one Landau level $\Lambda_{j}$ (see the figure below). We treat only the case $\lambda \in \sigma_{d} (H) \cap \mathbb{C}_{+}$. The same is true for $\lambda \in \sigma_{d} (H) \cap \mathbb{C}_{-}$ by considering rectangles $\Pi_{j}$ in the lower half plane $\mathbb{C}_{-}$. For simplicity in the sequel, by $\lambda_{0}$ we mean $\lambda_{0,j} \in \Pi_{j}$ and we assume that it satisfies condition \eqref{eq3.91}.
\begin{center}
\tikzstyle{ddEncadre}=[densely dotted]
\tikzstyle{grisEncadre}=[dashed]
\tikzstyle{grissEncadre}=[fill=gray!20]
\begin{tikzpicture}[scale = 0.8]
\draw (1.5,-1.5) -- (12,-1.5);
\node at (12.3,-1.5) {$\mathbb{R}$};
\draw [grissEncadre] (5,-1.5) -- (5,5) -- (9,5) -- (9,-1.5) -- cycle;
\draw [ddEncadre] (5,-1.5) -- (9,5);
\draw [ddEncadre] (5,5) -- (9,-1.5);
\node at (7.07,1.8) {$\lambda_{0}$};
\node at (7,3) {$\Pi_{j}$};

\node at (9.1,-1.8) {$\lambda_{4}$};
\node at (5,-1.8) {$\lambda_{1}$};
\node at (9.1,5.3) {$\lambda_{3}$};
\node at (5,5.3) {$\lambda_{2}$};

\node at (7.6,-1.2) {$\mathcal{L}_{1,\lambda}$};
\node at (7.1,-0.8) {\tiny{$S_{1}$}};

\node at (7.6,5.3) {$\mathcal{L}_{3,\lambda}$};
\node at (7.1,4.1) {\tiny{$S_{3}$}};

\node at (9.45,2.2) {$\mathcal{L}_{2,\lambda}$};
\node at (8.3,1.4) {\tiny{$S_{2}$}};

\node at (4.6,2.2) {$\mathcal{L}_{4,\lambda}$};
\node at (5.8,1.4) {\tiny{$S_{4}$}};

\node at (7,-1.52) {\tiny{$\bullet$}};
\node at (7,-1.85) {$\Lambda_{j}$};

\node at (11,-1.52) {\tiny{$\bullet$}};
\node at (11,-1.85) {$\Lambda_{j+1}$};

\node at (3,-1.52) {\tiny{$\bullet$}};
\node at (3,-1.85) {$\Lambda_{j-1}$};

\draw (2,-2) -- (2,6.5);
\draw [grisEncadre] (3,-1.52) -- (3,6);
\draw [grisEncadre] (11,-1.52) -- (11,6);

\draw [grisEncadre] (2,0.5) -- (11.7,0.5);;
\node at (1.4,0.5) {$\Vert V \Vert_{\infty}$};

\node at (4,-1.5) {\tiny{$||$}};
\node at (6,-1.5) {\tiny{$||$}};
\node at (8,-1.5) {\tiny{$||$}};
\node at (10,-1.5) {\tiny{$||$}};
\end{tikzpicture}

\vspace*{0.2cm}

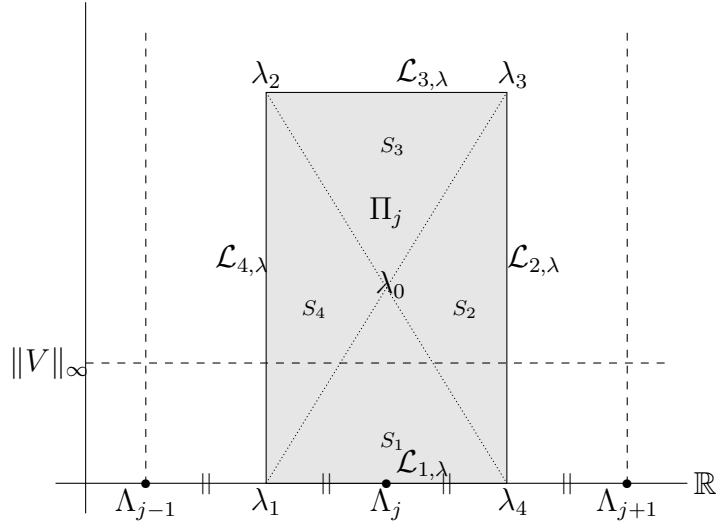
\textsc{Figure $2$.} Rectangle $\Pi_{j}$ in the upper half-plane $\mathbb{C}_{+}$ and containing $\Lambda_{j}$ as the only Landau levels.
\end{center}

We have $\varphi_{j}(1) = \Lambda_{j}$ and $\varphi_{j}(0) = \lambda_{0}$ where $\varphi_{j} : \mathbb{D} \rightarrow \Pi_{j}$ is the conformal map defined in subsection 4.2 with respect to the rectangle $\Pi_{j}$. The four triangles $S_{k}$, $1 \leq k \leq 4$ form a partition of the rectangle $\Pi_{j}$. Let $f$ be the function defined by $\eqref{eq3.131}$ and define $h_{j} : \mathbb{D} \rightarrow \mathbb{C}$ by 
\begin{equation}\label{eq3.181}
h_{j}(z) = f(\varphi_{j}(z)) \quad \textup{and} \quad \tilde{h}_{j} (z)= \frac{h_{j}(z)}{h_{j}(0)}.
\end{equation}
Then $\tilde{h}_{j}$ is holomorphic in the unit disk and $\eqref{eq3.14}$  implies that 
\begin{equation}\label{eq3.18}
\sigma_{d}(H) \cap \Pi_{j} = \lbrace \varphi_{j}(z) \in \Pi_{j} : z \in \mathbb{D} : \tilde{h}_{j}(z) = 0 \rbrace.
\end{equation}
We have the following lemma.

\begin{lem}\label{lem4}
Under the assumptions of Lemma $\ref{lem2}$, for any $z \in \mathbb{D}$,
\begin{equation}\label{eq3.19}
\log \vert \tilde{h}_{j}(z) \vert \leq \frac{C(p,b,d,j) K_{3}}{\textup{dist} \big(z,\mathbb{T}\big)^{\frac{p}{2}} \hspace{0.5mm} \vert z - 1 \vert^{\frac{p}{4}}},
\end{equation}
where the constant $C(p,b,d,j)$ satisfies the asymptotic property
\begin{equation}\label{eq3.20} 
C(p,b,d,j) \underset{j \rightarrow \infty} {\sim} j^{d+\frac{1}{2}},
\end{equation}
and the constant $K_{3}$ is defined by
\begin{equation}\label{eq1,5}
K_{3} := \Vert F \Vert_{L^{p}}^{p} \big{(} \Vert G \Vert_{L^{2}} + \Vert G \Vert_{L^{\infty}} \big{)}^{p} \big{(} 1 + \Vert V \Vert_{\infty} \big{)}^{d + \frac{1}{2}} .
\end{equation} 
\end{lem}

\hspace{-0.54cm} \begin{prof}
Constants are generic (changing from a relation to another). Let $K_{1}$ be the constant defined by \eqref{eq1,50}. For $\lambda$, $\lambda_{0} = \lambda_{0,j} \in \Pi_{j}$ with $\lambda_{0}$ satisfying condition \eqref{eq3.91}, using \eqref{eq1,00} we get the inequality
\begin{equation}\label{eq3.201} 
\begin{split}
& \small{\frac{\Gamma_{p} C(p,b,d) K_{1} (1 + \vert \lambda \vert)^{d+\frac{1}{2}}}{\textup{dist} \big(\lambda,[\Lambda_{0},+\infty)\big)^{\frac{p}{2}} \hspace{0.5mm} \textup{dist} (\lambda,E)^{\frac{p}{4}}} +  \Gamma_{p} C(p) K_{1} (1 + \vert \lambda_{0} \vert)^{d+\frac{1}{2}}} \\
& \small{\leq \frac{\Gamma_{p} C(p,b,d) K_{3} (1 + j)^{d+\frac{1}{2}}}{\textup{dist} \big(\lambda,\mathcal{L}_{1,\lambda} \big)^{\frac{p}{2}} \hspace{0.5mm} \vert \lambda - \Lambda_{j} \vert^{\frac{p}{4}}} + \Gamma_{p} C(p) K_{3} (1 + j)^{d + \frac{1}{2}}}.
\end{split}
\end{equation}
Since $\varphi_{j}(z) = \lambda$ and $\varphi_{j}(1) = \Lambda_{j}$, \eqref{eq3.17} (or Lemma \ref{lem3}) implies that
\begin{align*}
\small{\frac{\Gamma_{p} C(p,b,d) K_{3} (1 + j)^{d+\frac{1}{2}}}{\textup{dist} \big(\lambda,\mathcal{L}_{1,\lambda} \big)^{\frac{p}{2}} \hspace{0.5mm} \vert \lambda - \Lambda_{j} \vert^{\frac{p}{4}}}} & \small{\simeq \frac{C(p,d,j) K_{3} \hspace{0.5mm} \textup{dist} \big( z,\lbrace z_{1},z_{4} \rbrace \big)^{\frac{p}{4}}}{\textup{dist} \big(z,\mathbb{T} \big)^{\frac{p}{2}} \hspace{0.5mm} \vert z - 1 \vert^{\frac{p}{4}}}} \\
& \small{\leq \frac{C(p,d,j) K_{3}}{\textup{dist} \big(z,\mathbb{T} \big)^{\frac{p}{2}} \hspace{0.5mm} \vert z - 1 \vert^{\frac{p}{4}}}}.
\end{align*}
This together with \eqref{eq3.201} show that for any $\lambda \in \Pi_{j}$,
\begin{equation}\label{eq3.202} 
\begin{split}
& \small{\frac{\Gamma_{p} C(p,b,d) K_{1} (1 + \vert \lambda \vert)^{d+\frac{1}{2}}}{\textup{dist} \big(\lambda,[\Lambda_{0},+\infty)\big)^{\frac{p}{2}} \hspace{0.5mm} \textup{dist} (\lambda,E)^{\frac{p}{4}}} +  \Gamma_{p} C(p) K_{1} (1 + \vert \lambda_{0} \vert)^{d+\frac{1}{2}}} \\
& \small{\leq \frac{C(p,b,d,j) K_{3}}{\textup{dist} \big(z,\mathbb{T} \big)^{\frac{p}{2}} \hspace{0.5mm} \vert z - 1 \vert^{\frac{p}{4}}} + C(p,d,j) K_{3}} \\
& \small{\leq \frac{C(p,b,d,j) K_{3}}{\textup{dist} \big(z,\mathbb{T} \big)^{\frac{p}{2}} \hspace{0.5mm} \vert z - 1 \vert^{\frac{p}{4}}}}.
\end{split}
\end{equation}
Now \eqref{eq3.19} follows from Lemma \ref{lem2}, \eqref{eq3.181} and \eqref{eq3.202}.
\end{prof}

Applying Theorem \ref{theo2} to the function $\tilde{h}_{j}$ satisfying \eqref{eq3.19} in Lemma \ref{lem4}, we get for any $0 < \varepsilon < 1$
\begin{equation}\label{eq3.21} 
\sum_{\left\lbrace \tilde{h}_{j}(z)=0 \right\rbrace} \textup{dist} \big(z,\mathbb{T} \big)^{\frac{p}{2} + 1 + \varepsilon} \vert z - 1 \vert^{(\frac{p}{4} - 1 + \varepsilon)_{+}} \leq C(p,d,j,\varepsilon) K_{3}.
\end{equation}
Equivalently, using Lemma \ref{lem3}, estimate \eqref{eq3.21} can be written in $\Pi_{j}$ as 
\begin{equation}\label{eq3.22} 
\small{\sum_{\lambda \in \sigma_{d}(H) \cap \Pi_{j}} \Big{(} \textup{dist} \big(\lambda,\partial \Pi_{j}) \hspace{0.5mm} \textup{dist} \big(\lambda,\mathcal{F}_{\lambda}) \Big{)}^{\frac{p}{2} + 1 + \varepsilon} \vert \lambda - \Lambda_{j} \vert^{(\frac{p}{4} - 1 + \varepsilon)_{+}} \leq C(p,d,j,\varepsilon) K_{3}},
\end{equation}
where the constant $C(p,d,j,\varepsilon)$ satisfies again the asymptotic property \eqref{eq3.20} above. In the sequel we want to derive from \eqref{eq3.22} a bound of the quantity 
\begin{equation}\label{eq3.221} 
\sum_{\lambda \in \sigma_{d}(H) \cap \Pi_{j}} \textup{dist} \big( \lambda,[\Lambda_{0},+\infty) \big)^{\frac{p}{2} + 1 + \varepsilon} \hspace{0.5mm} \textup{dist} (\lambda,E)^{(\frac{p}{4} - 1 + \varepsilon)_{+}}.
\end{equation}
Note that in \eqref{eq3.22}, we can have accumulation of $\lambda \in \sigma_{d}(H)$ on the edges $\mathcal{L}_{2,\lambda}$ and $\mathcal{L}_{4,\lambda}$ of the boundary $\partial \Pi_{j}$ of $\Pi_{j}$. This is not due to the nature of the problem, but to the method we use. To treat this problem appearing in \eqref{eq3.22}, the idea is to deal for any rectangle $\Pi_{j}$ with its magnified version $\Pi'_{j}$ in the horizontal direction as in the figure below, where the constant $\delta$ is such that $0 < \delta < b$.

\begin{center}
\tikzstyle{ddEncadre}=[densely dotted]
\tikzstyle{grisEncadre}=[dashed]
\tikzstyle{grissEncadre}=[fill=gray!50]
\tikzstyle{grisssEncadre}=[fill=gray!20]
\begin{tikzpicture}[scale = 0.8]
\draw (1.5,-1.5) -- (14.3,-1.5);
\node at (14.5,-1.5) {$\mathbb{R}$};
\draw [grissEncadre] (5,-1.5) -- (5,5) -- (9,5) -- (9,-1.5) -- cycle;
\draw [grisssEncadre] (9,5) -- (9,-1.5) -- (13,-1.5) -- (13,5) -- cycle;

\draw [ddEncadre] (4.5,-1.5) -- (9.5,5);
\draw [ddEncadre] (4.5,5) -- (9.5,-1.5);

\node at (7.07,1.8) {$\lambda_{0}$};
\node at (7,3) {$\Pi_{j}$};

\node at (9.7,-1.8) {$\lambda'_{4}$};
\node at (4.3,-1.8) {$\lambda'_{1}$};
\node at (9.7,5.3) {$\lambda'_{3}$};
\node at (4.3,5.3) {$\lambda'_{2}$};

\node at (7.8,-1.2) {\tiny{$\mathcal{L}'_{1,\lambda}$}};
\node at (7.1,-0.8) {\tiny{$S'_{1}$}};

\node at (7.8,5.2) {\tiny{$\mathcal{L}'_{3,\lambda}$}};
\node at (7.1,4.1) {\tiny{$S'_{3}$}};

\node at (9.9,2.2) {\tiny{$\mathcal{L}'_{2,\lambda}$}};
\node at (8.3,1.4) {\tiny{$S'_{2}$}};

\node at (4.15,2.2) {\tiny{$\mathcal{L}'_{4,\lambda}$}};
\node at (5.8,1.4) {\tiny{$S'_{4}$}};

\node at (7,-1.52) {\tiny{$\bullet$}};
\node at (7,-1.85) {$\Lambda_{j}$};

\node at (3,-1.52) {\tiny{$\bullet$}};
\node at (3,-1.85) {$\Lambda_{j-1}$};

\draw (2,-3.3) -- (2,7);

\node at (11,-1.52) {\tiny{$\bullet$}};
\node at (11,-1.85) {$\Lambda_{j+1}$};

\node at (12,1.2) {$\Pi_{j+1}$};

\draw (4.5,-1.5) -- (4.5,5) -- (9.5,5) -- (9.5,-1.5) -- cycle;

\node at (7.1,6.44) {$\Pi'_{j}$};
\draw [->,>=latex] (7.3,6.5) to[bend right = 0] (9.5,6.5);
\draw [->,>=latex] (6.8,6.5) to[bend right = 0] (4.5,6.5);

\draw [grisEncadre] (8.5,5) -- (8.5,-1.53);
\draw [grisEncadre] (13.5,5) -- (13.5,-1.53);
\draw (13,5) -- (13.5,5);

\node at (11.1,-2.8) {$\Pi'_{j+1}$};
\draw [grisEncadre] [->,>=latex] (11.5,-2.7) to[bend right = 0] (13.5,-2.7);
\draw [grisEncadre] [->,>=latex] (10.6,-2.7) to[bend right = 0] (8.5,-2.7);

\draw [<->,>=latex] (4.5,-1.2) to[bend right = 0] (5,-1.2);
\node at (4.75,-1) {\tiny{$\delta$}};

\draw [<->,>=latex] (9,-1.2) to[bend right = 0] (9.5,-1.2);
\node at (9.25,-1) {\tiny{$\delta$}};

\draw [<->,>=latex] (13,-1.2) to[bend right = 0] (13.5,-1.2);
\node at (13.25,-1) {\tiny{$\delta$}};

\draw [grisEncadre] (2,0.5) -- (14,0.5);;
\node at (1.4,0.5) {$\Vert V \Vert_{\infty}$};

\node at (4,-1.5) {\tiny{$||$}};
\node at (6,-1.5) {\tiny{$||$}};
\node at (8,-1.5) {\tiny{$||$}};
\node at (10,-1.5) {\tiny{$||$}};
\node at (12,-1.5) {\tiny{$||$}};
\end{tikzpicture}

\vspace*{0.2cm}

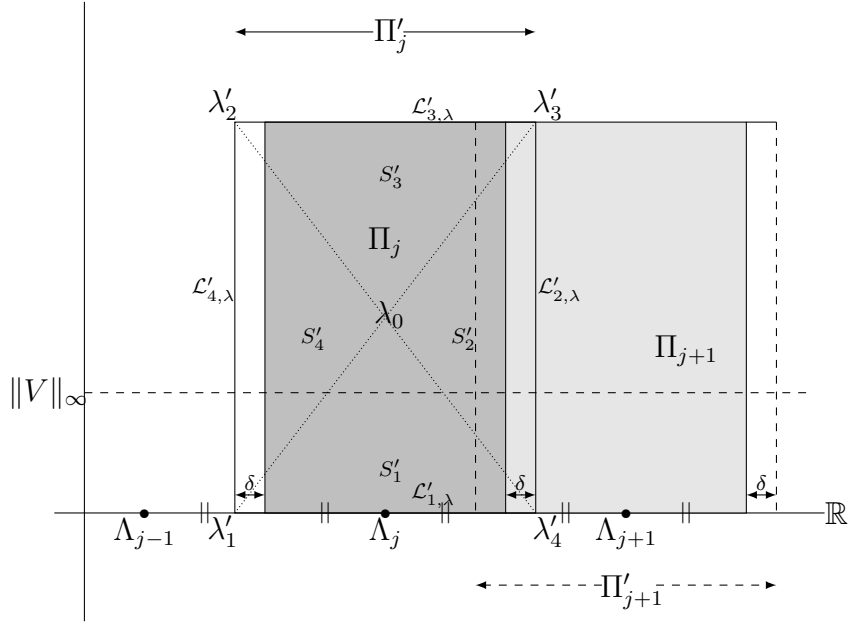
\textsc{Figure $3$.} Magnified versions $\Pi'_{j}$ and $\Pi'_{j+1}$ respectively of disjoint rectangles $\Pi_{j}$ and $\Pi_{j+1}$ in the upper half-plane $\mathbb{C}_{+}$.
\end{center}
In the figure above, we introduce a partition of the rectangle $\Pi'_{j}$ by the four triangles $S'_{k}$, $1 \leq k \leq 4$. Applying \eqref{eq3.22} to the rectangle $\Pi'_{j}$, we get
$$\small{\sum_{\lambda \in \sigma_{d}(H) \cap \Pi'_{j}} \Big{(} \textup{dist} \big(\lambda,\partial \Pi'_{j}\big) \hspace{0.5mm} \textup{dist} \big(\lambda,\mathcal{F}_{\lambda'}\big) \Big{)}^{\frac{p}{2} + 1 + \varepsilon} \vert \lambda - \Lambda_{j} \vert^{(\frac{p}{4} - 1 + \varepsilon)_{+}} \leq C(p,b,d,j,\varepsilon) K_{3}}.$$
Since $\Pi_{j} \subset \Pi'_{j}$, then the sum taken on $\Pi_{j}$ gives
\begin{equation}\label{eq3.31} 
\small{\sum_{\lambda \in \sigma_{d}(H) \cap \Pi_{j}} \Big{(} \textup{dist} \big(\lambda,\partial \Pi'_{j}\big) \hspace{0.5mm} \textup{dist} \big(\lambda,\mathcal{F}_{\lambda'}\big) \Big{)}^{\frac{p}{2} + 1 + \varepsilon} \vert \lambda - \Lambda_{j} \vert^{(\frac{p}{4} - 1 + \varepsilon)_{+}} \leq C(p,b,d,j,\varepsilon) K_{3}}.
\end{equation}
Otherwise, since there are no eigenvalues in the sector $S'_{3}$, we have
\begin{equation}\label{eq3.311} 
\begin{split}
\small{\sum_{\lambda \in \sigma_{d}(H) \cap \Pi_{j}}} & \small{\Big{(} \textup{dist} \big(\lambda,\partial \Pi'_{j}\big) \hspace{0.5mm} \textup{dist} \big(\lambda,\mathcal{F}_{\lambda'}\big) \Big{)}^{\frac{p}{2} + 1 + \varepsilon} \vert \lambda - \Lambda_{j} \vert^{(\frac{p}{4} - 1 + \varepsilon)_{+}} =} \\
& \small{\sum_{\lambda \in \sigma_{d}(H) \cap \Pi_{j} \cap S'_{1}} \Big{(} \textup{dist} \big(\lambda,\mathcal{L}'_{1,\lambda}\big) \hspace{0.5mm} \textup{dist} \big( \lambda,\lbrace \lambda'_{1},\lambda'_{4} \rbrace \big) \Big{)}^{\frac{p}{2} + 1 + \varepsilon} \vert \lambda - \Lambda_{j} \vert^{(\frac{p}{4} - 1 + \varepsilon)_{+}}} \\
& \small{+ \sum_{\lambda \in \sigma_{d}(H) \cap \Pi_{j} \cap S'_{2}} \Big{(} \textup{dist} \big(\lambda,\mathcal{L}'_{2,\lambda}\big) \hspace{0.5mm} \textup{dist} \big( \lambda,\lbrace \lambda'_{4},\lambda'_{3} \rbrace \big) \Big{)}^{\frac{p}{2} + 1 + \varepsilon} \vert \lambda - \Lambda_{j} \vert^{(\frac{p}{4} - 1 + \varepsilon)_{+}}} \\
& \small{+ \sum_{\lambda \in \sigma_{d}(H) \cap \Pi_{j} \cap S'_{4}} \Big{(} \textup{dist} \big(\lambda,\mathcal{L}'_{4,\lambda}\big) \hspace{0.5mm} \textup{dist} \big( \lambda,\lbrace \lambda'_{2},\lambda'_{1} \rbrace \big) \Big{)}^{\frac{p}{2} + 1 + \varepsilon} \vert \lambda - \Lambda_{j} \vert^{(\frac{p}{4} - 1 + \varepsilon)_{+}}}.
\end{split}
\end{equation}
This together with \eqref{eq3.31} implies in particular that
\begin{equation}\label{eq3.32} 
\begin{split}
\sum_{\lambda \in \sigma_{d}(H) \cap \Pi_{j} \cap S'_{2}} & \Big{(} \textup{dist} \big(\lambda,\mathcal{L}'_{2,\lambda}\big) \hspace{0.5mm} \textup{dist} \big( \lambda,\lbrace \lambda'_{4},\lambda'_{3} \rbrace \big) \Big{)}^{\frac{p}{2} + 1 + \varepsilon} \vert \lambda - \Lambda_{j} \vert^{(\frac{p}{4} - 1 + \varepsilon)_{+}} \\
& \leq C(p,b,d,j,\varepsilon) K_{3}, \\
\sum_{\lambda \in \sigma_{d}(H) \cap \Pi_{j} \cap S'_{4}} & \Big{(} \textup{dist} \big(\lambda,\mathcal{L}'_{4,\lambda}\big) \hspace{0.5mm} \textup{dist} \big( \lambda,\lbrace \lambda'_{2},\lambda'_{1} \rbrace \big) \Big{)}^{\frac{p}{2} + 1 + \varepsilon} \vert \lambda - \Lambda_{j} \vert^{(\frac{p}{4} - 1 + \varepsilon)_{+}} \\
& \leq C(p,b,d,j,\varepsilon) K_{3}.
\end{split}
\end{equation}
Take into account the fact that there are no eigenvalues in the sector $S'_{3}$, clearly quantity \eqref{eq3.221} above can be rewritten as
\begin{equation}\label{eq3.33} 
\begin{split}
\small{\sum_{\lambda \in \sigma_{d}(H) \cap \Pi_{j}}} & \small{\textup{dist} \big(\lambda,\mathcal{L}_{1,\lambda}) ^{\frac{p}{2} + 1 + \varepsilon} \vert \lambda - \Lambda_{j} \vert^{(\frac{p}{4} - 1 + \varepsilon)_{+}} =} \\
& \small{\sum_{\lambda \in \sigma_{d}(H) \cap \Pi_{j} \cap S'_{1}} \textup{dist} \big(\lambda,\mathcal{L}_{1,\lambda})^{\frac{p}{2} + 1 + \varepsilon} \vert \lambda - \Lambda_{j} \vert^{(\frac{p}{4} - 1 + \varepsilon)_{+}}} \\
& \small{+ \sum_{\lambda \in \sigma_{d}(H) \cap \Pi_{j} \cap S'_{2}} \textup{dist} \big(\lambda,\mathcal{L}_{1,\lambda})^{\frac{p}{2} + 1 + \varepsilon} \vert \lambda - \Lambda_{j} \vert^{(\frac{p}{4} - 1 + \varepsilon)_{+}}} \\
& \small{+ \sum_{\lambda \in \sigma_{d}(H) \cap \Pi_{j} \cap S'_{4}} \textup{dist} \big(\lambda,\mathcal{L}_{1,\lambda})^{\frac{p}{2} + 1 + \varepsilon} \vert \lambda - \Lambda_{j} \vert^{(\frac{p}{4} - 1 + \varepsilon)_{+}}}.
\end{split}
\end{equation}
By combining \eqref{eq3.31} and \eqref{eq3.311}, and using the first term of RHS in \eqref{eq3.311} and the lower bound $\textup{dist} \big( \lambda,\lbrace \lambda'_{1},\lambda'_{4} \rbrace \big) \geq \delta$, we get the following bound
\begin{equation}\label{eq3.331} 
\small{\sum_{\lambda \in \sigma_{d}(H) \cap \Pi_{j} \cap S'_{1}} \textup{dist} \big(\lambda,\mathcal{L}_{1,\lambda}) ^{\frac{p}{2} + 1 + \varepsilon} \vert \lambda - \Lambda_{j} \vert^{(\frac{p}{4} - 1 + \varepsilon)_{+}} \leq \frac{C(p,b,d,j,\varepsilon) K_{3}}{\delta^{\frac{p}{2} + 1 + \varepsilon}}}.
\end{equation}
For $\lambda \in \sigma_{d}(H) \cap \Pi_{j} \cap S'_{2}$, it can be easily checked that $\textup{dist} \big(\lambda,\mathcal{L}_{1,\lambda}) \leq \Vert V \Vert_{\infty}$ and $\textup{dist} \big(\lambda,\mathcal{L}'_{2,\lambda}\big) \hspace{0.5mm} \textup{dist} \big( \lambda,\lbrace \lambda'_{4},\lambda'_{3} \rbrace \big) \geq \delta^{2}$. Thus
\begin{align*}
& \small{\sum_{\lambda \in \sigma_{d}(H) \cap \Pi_{j} \cap S'_{2}} \textup{dist} \big(\lambda,\mathcal{L}_{1,\lambda})^{\frac{p}{2} + 1 + \varepsilon} \vert \lambda - \Lambda_{j} \vert^{(\frac{p}{4} - 1 + \varepsilon)_{+}}} \\
& \small{\leq \frac{\Vert V \Vert_{\infty}^{\frac{p}{2} + 1 + \varepsilon}}{\delta^{2(\frac{p}{2} + 1 + \varepsilon)}} \sum_{\lambda \in \sigma_{d}(H) \cap \Pi_{j} \cap S'_{2}} \Big{(} \textup{dist} \big(\lambda,\mathcal{L}'_{2,\lambda}\big) \hspace{0.5mm} \textup{dist} \big( \lambda,\lbrace \lambda'_{4},\lambda'_{3} \rbrace \big) \Big{)}^{\frac{p}{2} + 1 + \varepsilon} \vert \lambda - \Lambda_{j} \vert^{(\frac{p}{4} - 1 + \varepsilon)_{+}}}.
\end{align*}
Now using \eqref{eq3.32} we get the bound
\begin{equation}\label{eq3.332} 
\small{\sum_{\lambda \in \sigma_{d}(H) \cap \Pi_{j} \cap S'_{2}} \textup{dist} \big(\lambda,\mathcal{L}_{1,\lambda})^{\frac{p}{2} + 1 + \varepsilon} \vert \lambda - \Lambda_{j} \vert^{(\frac{p}{4} - 1 + \varepsilon)_{+}} \leq C(p,b,d,j,\varepsilon) K} 
\end{equation} 
where the constant $K$ is defined by \eqref{eq1,005}. By similar arguments, we show that
\begin{equation}\label{eq3.333} 
\small{\sum_{\lambda \in \sigma_{d}(H) \cap \Pi_{j} \cap S'_{4}} \textup{dist} \big(\lambda,\mathcal{L}_{1,\lambda})^{\frac{p}{2} + 1 + \varepsilon} \vert \lambda - \Lambda_{j} \vert^{(\frac{p}{4} - 1 + \varepsilon)_{+}} \leq C(p,b,d,j,\varepsilon) K} 
\end{equation} 
For the rectangle $\Pi_{0}$ containing the first Landau level $\Lambda_{0}$, choose the vertice $\lambda_{1}$ of the edge $\mathcal{L}_{4,\lambda}$ so that $\lambda_{1} \leq - \Vert V \Vert_{\infty}$. Then estimates \eqref{eq3.33}-\eqref{eq3.333} imply that for any $j \geq 0$
\begin{equation}\label{eq3.351} 
\small{\sum_{\lambda \in \sigma_{d}(H) \cap \Pi_{j}} \textup{dist} \big(\lambda,\mathcal{L}_{1,\lambda})^{\frac{p}{2} + 1 + \varepsilon} \vert \lambda - \Lambda_{j} \vert^{(\frac{p}{4} - 1 + \varepsilon)_{+}} \leq C(p,b,d,j,\varepsilon) K}.
\end{equation}
Thus we have proved the following proposition.

\begin{prop}\label{pro1}
For any $j \geq 0$ the following bound
\begin{equation}\label{eq3.36} 
\small{\sum_{\lambda \in \sigma_{d}(H) \cap \Pi_{j}} \textup{dist} \big( \lambda,[\Lambda_{0},+\infty) \big)^{\frac{p}{2} + 1 + \varepsilon} \hspace{0.5mm} \textup{dist} (\lambda,E)^{(\frac{p}{4} - 1 + \varepsilon)_{+}} \leq C(p,b,d,j,\varepsilon) K}
\end{equation}
holds, where $K$ is defined by \eqref{eq1,005} and the constant $C(p,b,d,j,\varepsilon)$ satisfies the asymptotic property \eqref{eq3.20}.
\end{prop}

Now we go back to the global bound.

\subsection{Proof of Theorem \ref{theo1}}

The main idea is to do with the help of \eqref{eq3.36} a summation on the index $j$. The only way to obtain a finite sum with respect to $j$ is first to multiply \eqref{eq3.36} by an appropriate weight function of $j$, taking into account the asymptotic property \eqref{eq3.20} of the constant $C(p,b,d,j,\varepsilon)$, namely
$$C(p,b,d,j,\varepsilon) \underset{j \rightarrow \infty} {\sim} j^{d+\frac{1}{2}}.$$
Let $\gamma$ be such that $\gamma - (d + \frac{1}{2}) > 1$ or equivalently $\gamma > d + \frac{3}{2}$. By \eqref{eq3.36}, we have
\begin{equation}\label{eq3.38} 
\begin{split}
\small{\frac{1}{(1 + j)^{\gamma}} \sum_{\lambda \in \sigma_{d}(H) \cap \Pi_{j}}} & \small{\textup{dist} \big( \lambda,[\Lambda_{0},+\infty) \big)^{\frac{p}{2} + 1 + \varepsilon} \hspace{0.5mm} \textup{dist} (\lambda,E)^{(\frac{p}{4} - 1 + \varepsilon)_{+}}} \\
& \small{\leq \frac{C(p,b,d,j,\varepsilon)}{(1 + j)^{\gamma}} K}.
\end{split}
\end{equation}
Taking in \eqref{eq3.38} the sum with respect to $j$, we get
\begin{equation}\label{eq3.39} 
\begin{split}
\small{\sum_{j} \hspace*{0.2cm} \frac{1}{(1 + j)^{\gamma}} \sum_{\lambda \in \sigma_{d}(H) \cap \Pi_{j}}} & \small{\textup{dist} \big( \lambda,[\Lambda_{0},+\infty) \big)^{\frac{p}{2} + 1 + \varepsilon} \hspace{0.5mm} \textup{dist} (\lambda,E)^{(\frac{p}{4} - 1 + \varepsilon)_{+}}} \\
& \small{\leq \hspace*{0.2cm} \sum_{j} \frac{C(p,b,d,j,\varepsilon)}{(1 + j)^{\gamma}} K}.
\end{split}
\end{equation}
By the above choice of $\gamma$, RHS in \eqref{eq3.39} is convergent so that
\begin{equation}\label{eq3.40} 
\small{\sum_{j} \frac{C(p,b,d,j,\varepsilon)}{(1 + j)^{\gamma}} K = C(p,b,d,\varepsilon) K}.
\end{equation}
Thus using the fact that for any $\lambda \in \Pi_{j}$ we have $1 + j \simeq 1 + \vert \lambda \vert$, we get
\begin{equation}\label{eq3.42} 
\small{\sum_{\lambda \in \sigma_{d}(H) \cap \mathbb{C_{+}}} \frac{\textup{dist} \big( \lambda,[\Lambda_{0},+\infty) \big)^{\frac{p}{2} + 1 + \varepsilon} \hspace{0.5mm} \textup{dist} (\lambda,E)^{(\frac{p}{4} - 1 + \varepsilon)_{+}}}{(1 + \vert \lambda \vert)^{\gamma}} \leq C_{0} K}
\end{equation}
where $C_{0} = C(p,b,d,\varepsilon)$. Since \eqref{eq3.42} is true for $\lambda \in \sigma_{d}(H) \cap \mathbb{C_{-}}$ by considering rectangles $\Pi_{j}$ in the lower half plane $\mathbb{C}_{-}$, then finally we have
\begin{equation}\label{eq3.43} 
\small{\sum_{\lambda \in \sigma_{d}(H)} \frac{\textup{dist} \big( \lambda,[\Lambda_{0},+\infty) \big)^{\frac{p}{2} + 1 + \varepsilon} \hspace{0.5mm} \textup{dist} (\lambda,E)^{(\frac{p}{4} - 1 + \varepsilon)_{+}}}{(1 + \vert \lambda \vert)^{\gamma}} \leq C_{0} K}.
\end{equation}
This concludes the proof of Theorem 2.1.

\section{About the dimension $2d$, $d \geq 1$}

Consider the Schrödinger operator $H_{0,\perp}$ defined by \eqref{eq0,1}. In this section, we investigate the discrete spectrum $\sigma_{d}(H_{\perp})$ of the perturbed operator $H_{\perp}$ defined by \eqref{eq0.3}, and we assume that the electric potential $V$ satisfies assumption \eqref{eq1,1} for $n = 2d$.

\subsection{Estimate of the resolvent}

Recall that the spectrum $\sigma(H_{0,\perp})$ of $H_{0,\perp}$ is a discrete set and consists of the Landau levels $\Lambda_{j}$ defined by \eqref{eq1,00}, $j \in \mathbb{N}$. So for any $\lambda \in \mathbb{C} \setminus \sigma(H_{0,\perp})$ the resolvent of $H_{0,\perp}$ is given by
\begin{equation}\label{eq0,6}
\displaystyle (H_{0,\perp} - \lambda)^{-1} = \sum_{j \in \mathbb{N}}(\Lambda_{j} - \lambda)^{-1} p_{j} ,
\end{equation}
where $p_{j}$ is the orthogonal projection onto $\ker \hspace{0.5mm} (H_{0,\perp} - \Lambda_{j})$. 

\begin{lem}\label{lem01}
Let $n = 2d$, $d \geq 1$ and $\lambda \in \mathbb{C} \setminus E$ where $E$ is the set of Landau levels defined by \eqref{eq1,00}. Assume that $F \in L^{p}\big(\mathbb{R}^{n}\big)$ with $p \geq 2 \big[ \frac{d}{2} \big] + 2$. Then there exists a constant $C = C(p)$ such that
\begin{equation}\label{est004}
\left\Vert F (H_{0,\perp} - \lambda )^{-1} \right\Vert_{p}^{p} \leq \frac{C (1 + \vert \lambda \vert)^{d} \Vert F \Vert_{L^{p}}^{p}}{\textup{dist} (\lambda,E)^{p}}.
\end{equation}
\end{lem}

\hspace{-0.54cm} \begin{prof}
Let $\lambda \in \mathbb{C} \setminus E$. We first show that \eqref{est004} holds if $p$ is even. Using the identity
$$(H_{0,\perp} - \lambda)^{-1} = \bigl( H_{0,\perp} + 1 + \vert \lambda \vert \bigr)^{-1} \bigl( (1 + \vert \lambda \vert + \lambda) (H_{0,\perp} - \lambda)^{-1} + I \bigr),$$ 
we get
\begin{equation}\label{eq0,7}
\begin{split}
\left\Vert F (H_{0,\perp} - \lambda)^{-1} \right\Vert^{p}_{p} \leq & \left\Vert F \bigl( H_{0,\perp} + 1 + \vert \lambda \vert \bigr)^{-1} \right\Vert^{p}_{p} \\
& \times \left\Vert \Bigl( (1 + \vert \lambda \vert + \lambda) (H_{0,\perp} - \lambda)^{-1} + I \Bigr) \right\Vert^{p}.
\end{split}
\end{equation}
Hence if $p$ is even, \eqref{est004} follows as in the proof of Lemma \ref{lem1} from \eqref{eq3.1} to \eqref{eq3.31b}, where the operator $H_{0}$ is replaced by $H_{0,\perp}$, $D_{x}^{2}$ is removed and $G = I$. 

To prove that \eqref{est004} holds for any $p \geq 2 \big[ \frac{d}{2} \big] + 2$, we interpolate by the Riesz-Thorin Theorem as in (ii) of the proof of Lemma \ref{lem1}. This concludes the proof.
\end{prof}

\subsection{Proof of Theorem \ref{theo3}}

Constants are generic (changing from a relation to another). The first important tool of the proof is the following result of Hansmann \cite[Theorem 1]{han}. Let $A_{0} = A_{0}^{\ast}$ be a bounded self-adjoint operator on a Hilbert space, $A$ a bounded operator such that $A - A_{0} \in S_{p}$, $p > 1$. Then 
\begin{equation}\label{eq0,8}
\sum_{\lambda \in \sigma_{d}(A)} \textup{dist} \big( \lambda,\sigma(A_{0})\big)^{p} \leq C \Vert A - A_{0} \Vert_{p}^{p},
\end{equation}
where $C$ is an explicit constant which depends only on $p$.

Now let us fix a constant $\mu$ defined by 
\begin{equation}\label{eq0,9}
\mu := - \Vert V \Vert_{\infty} - 1.
\end{equation}
Since $H_{0,\perp}$ and $H_{\perp}$ are not bounded operators, to apply \eqref{eq0,8} we shall consider the bounded resolvents 
\begin{equation}\label{eq0,10}
A_{0}(\mu) := (H_{0,\perp} - \mu)^{-1} \quad \text{and} \quad A(\mu) := (H_{\perp} - \mu)^{-1}.
\end{equation}
The assumption \eqref{eq1,1} on $V$ for $n = 2d$ implies that there exists a bounded operator $\mathcal{V}$ such that $V(X_{\perp}) = \mathcal{V} F(X_{\perp})$ for any $X_{\perp} \in \mathbb{R}^{n}$. So using the resolvent equation 
$$(H_{\perp} - \mu)^{-1} - (H_{0,\perp} - \mu)^{-1} = - (H_{\perp} - \mu)^{-1} V (H_{0,\perp} - \mu)^{-1},$$
we get
\begin{equation}\label{eq0,11}
\big\Vert A(\mu) - A_{0}(\mu) \big\Vert_{p}^{p} \leq C \big\Vert (H_{\perp} - \mu)^{-1} \big\Vert^{p} \big\Vert F(H_{0,\perp} - \mu)^{-1} \big\Vert_{p}^{p},
\end{equation}
with $C > 0$ a constant. The choice of the constant $\mu$ and \eqref{eq0.03} imply that $\textup{dist} \big(\mu,\overline{N(H_{\perp})}\big) \geq 1$, and by \cite[Lemma 9.3.14]{dav}
\begin{equation}\label{eq0,12}
\left\Vert (H_{\perp} - \mu)^{-1} \right\Vert \leq \frac{1}{\textup{dist} \big(\mu,\overline{N(H_{\perp})}\big)} \leq 1.
\end{equation}
Lemma \ref{lem01} and the choice of $\mu$ imply that 
\begin{equation}\label{eq0,13}
\Vert F(H_{0,\perp} - \mu)^{-1} \big\Vert_{p}^{p} \leq C \Vert F \Vert_{L^{p}}^{p}.
\end{equation}
By combining \eqref{eq0,11}, \eqref{eq0,12} and \eqref{eq0,13}, we finally get
\begin{equation}\label{eq0,14}
\big\Vert A(\mu) - A_{0}(\mu) \big\Vert_{p}^{p} \leq C \Vert F \Vert_{L^{p}}^{p}.
\end{equation}
Hence by applying \eqref{eq0,8} to the resolvents $A(\mu)$ and $A_{0}(\mu)$, we get
\begin{equation}\label{eq0,15}
\sum_{z \in \sigma_{d}(A(\mu))} \textup{dist} \big(z,\sigma(A_{0}(\mu))\big)^{p} \leq C \Vert F \Vert_{L^{p}}^{p},
\end{equation}
where $C = C(p)$. Let $z = \varphi_{\mu}(\lambda) = (\lambda - \mu)^{-1}$. The Spectral Mapping Theorem implies that
\begin{equation}\label{eq0,151}
z \in \sigma_{d} \big{(} A(\mu) \big{)} \quad \Big{(} z \in \sigma \big{(} A_{0}(\mu) \big{)} \Big{)} \quad \Longleftrightarrow \quad \lambda \in \sigma_{d}(H_{\perp}) \quad \Big{(} \lambda \in \sigma (H_{0,\perp}) \Big{)}.
\end{equation}
The second ingredient of the proof of the theorem is the following distortion lemma for the transformation $z = \varphi_{\mu}(\lambda) = (\lambda - \mu)^{-1}$.

\begin{lem}\label{lem02}
Let $\mu$ be the constant defined by \eqref{eq0,9} and $E$ be the set of Landau levels defined by \eqref{eq1,00}. Then the following bound holds
\begin{equation}\label{eq0,16}
\textup{dist} \big( \varphi_{\mu}(\lambda),\varphi_{\mu}(E) \big) \geq \frac{C \hspace{0.08cm} \textup{dist} \big(\lambda,E\big)}{\big( 1 + \Vert V \Vert_{\infty} \big)^{2} \big( 1 + \vert \lambda \vert \big)^{2}}, \quad \lambda \in \mathbb{C},
\end{equation}
where $C = C(b,d)$ is a constant depending on $b$ and $d$.
\end{lem}

The proof of Lemma \ref{lem02} follows directly from Lemma \ref{lem03} and Lemma \ref{lem04} below. For more comprehension in the sequel, it is convenient to give the figure which represents the transformation of the complex plane by the conformal map $\varphi_{\mu}$ (see the figure below). 

\begin{center}
\tikzstyle{ddEncadre}=[densely dotted]
\tikzstyle{grisEncadre}=[dashed]
\tikzstyle{grissEncadre}=[fill=gray!60]
\begin{tikzpicture}

\begin{scope}

\draw (0,-1.5) -- (2.5,-1.5);
\draw [very thick] (2.5,-1.5) -- (4.7,-1.5);
\draw (1,-4) -- (1,2.5);
\draw (2.5,-4) -- (2.5,2.5);

\node at (2.5,-1.52) {\tiny{$\bullet$}};
\node at (2.5,-1.75) {\tiny{$\Lambda_{0}$}};

\draw (1.75,-1.5) circle (0.75);

\node at (3,-1.52) {\tiny{$\bullet$}};
\node at (3.5,-1.52) {\tiny{$\bullet$}};
\node at (4,-1.52) {\tiny{$\bullet$}};
\node at (4.5,-1.52) {\tiny{$\bullet$}};

\draw [ddEncadre] (2.5,-1.52) -- (3.7,-1);
\draw [ddEncadre] (3,-1.52) -- (3.7,-1);
\draw [ddEncadre] (3.5,-1.52) -- (3.7,-1);
\draw [ddEncadre] (4,-1.52) -- (3.7,-1);
\draw [ddEncadre] (4.5,-1.52) -- (3.7,-1);

\node at (3.7,-0.83) {\tiny{$E$}};
\node at (1.75,-1.52) {\tiny{$\times$}};

\draw [->] (1.75,0.5) -- (1.75,-1.4);
\node at (1.75,0.8) {\tiny{$\frac{\mu+\Lambda_{0}}{2}$}};

\node at (1,-1.52) {\tiny{$\bullet$}};
\node at (0.8,-1.75) {\tiny{$\mu$}};

\node at (4.8,1.5) {\tiny{$\lambda \longmapsto \varphi_{\mu}(\lambda) = z = \frac{1}{\lambda-\mu}$}};
\node at (4,-2) {\tiny{$I = [\Lambda_{0},+\infty)$}};

\end{scope}

\begin{scope}[xshift=6cm]

\draw [grissEncadre] (1.75,-1.5) circle (0.75);

\draw (0,-1.5) -- (1,-1.5);
\draw [very thick] (1,-1.5) -- (2.5,-1.5);
\draw (2.5,-1.5) -- (4.7,-1.5);
\draw (1,-4) -- (1,2.5);
\draw (2.5,-4) -- (2.5,2.5);

\node at (2.5,-1.52) {\tiny{$\bullet$}};
\node at (2.8,-1.75) {\tiny{$\frac{1}{\Lambda_{0}-\mu}$}};

\node at (2.1,-1.52) {\tiny{$\bullet$}};
\node at (1.8,-1.52) {\tiny{$\bullet$}};
\node at (1.6,-1.52) {\tiny{$\bullet$}};
\node at (1.45,-1.52) {\tiny{$\bullet$}};

\draw [ddEncadre] (1,-1.52) -- (1.7,-0.3);
\draw [ddEncadre] (1.45,-1.52) -- (1.7,-0.3);
\draw [ddEncadre] (1.6,-1.52) -- (1.7,-0.3);
\draw [ddEncadre] (1.8,-1.52) -- (1.7,-0.3);
\draw [ddEncadre] (2.1,-1.52) -- (1.7,-0.3);
\draw [ddEncadre] (2.5,-1.52) -- (1.7,-0.3);

\node at (1.73,-0.15) {\tiny{$\varphi_{\mu}(E)$}};

\node at (1,-1.52) {\tiny{$\bullet$}};
\node at (0.8,-1.75) {\tiny{$0$}};

\node at (1.7,-3) {\tiny{$\varphi_{\mu}(I) = \Big[0,\frac{1}{\Lambda_{0}-\mu}\Big]$}};

\end{scope}
\end{tikzpicture}

\vspace*{0.2cm}

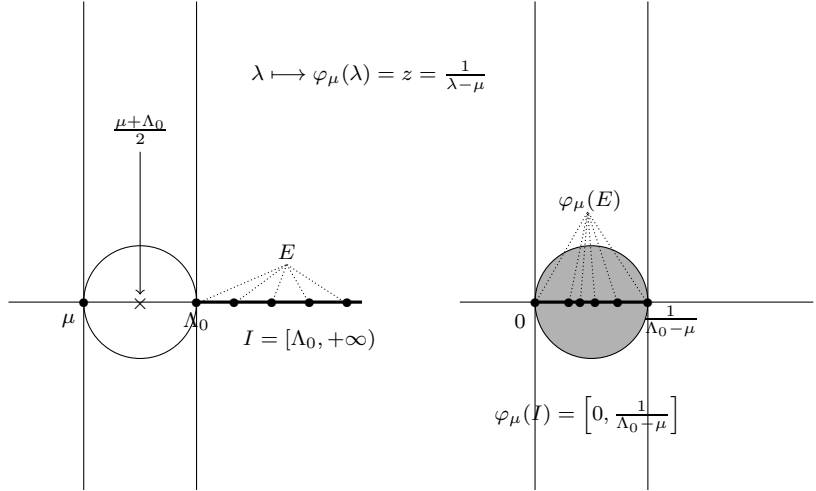
\textsc{Figure $4$.} Transformation of the complex plane by the conformal map $z = \varphi_{\mu}(\lambda) = \frac{1}{\lambda-\mu}$.
\end{center}

It can be easily checked that
\begin{equation}\label{eq0,17}
\small{\varphi_{\mu}\Big( \big\lbrace \lambda \in \mathbb{C} : \textup{Re} \hspace{0.5mm} \lambda < \mu \big\rbrace \Big) = \big\lbrace z \in \mathbb{C} : \textup{Re} \hspace{0.5mm} z < 0 \big\rbrace},
\end{equation}
\begin{equation}\label{eq0,18}
\small{\varphi_{\mu} \left( \left\lbrace \lambda \in \mathbb{C} : \left\vert \lambda - \frac{\mu+\Lambda_{0}}{2} \right\vert \leq \frac{\Lambda_{0}-\mu}{2} \right\rbrace \right) = \left\lbrace z \in \mathbb{C} : \textup{Re} \hspace{0.5mm} z \geq \frac{1}{\Lambda_{0}-\mu} \right\rbrace},
\end{equation}
\begin{equation}\label{eq0,19}
\small{\varphi_{\mu}\Big( \big\lbrace \mu \leq \textup{Re} \hspace{0.5mm} \lambda \leq \Lambda_{0} \big\rbrace \Big) = \big\lbrace z \in \mathbb{C} : \textup{Re} \hspace{0.5mm} z \geq 0  \big\rbrace \cap \text{outside the gray disk}},
\end{equation}
\begin{equation}\label{eq0,20}
\small{\varphi_{\mu} \Big( \big\lbrace \lambda \in \mathbb{C} : \textup{Re} \hspace{0.5mm} \lambda \geq \Lambda_{0} \big\rbrace \Big) = \left\lbrace z : \left\vert z - \frac{1}{2(\Lambda_{0}-\mu)} \right\vert \leq \frac{1}{2(\Lambda_{0}-\mu)} \right\rbrace}.
\end{equation}

\begin{lem}\label{lem03}
Let $I = [\Lambda_{0},+\infty)$. The following bound holds for any $\lambda \in \mathbb{C}$
\begin{equation}\label{eq0,21}
\textup{dist} \big(\varphi_{\mu}(\lambda),\varphi_{\mu}(I)\big) \geq \frac{C \hspace{0.08cm} \textup{dist} \big(\lambda,I \big)}{\big( 1 + \Vert V \Vert_{\infty} \big)^{2} \big( 1 + \vert \lambda \vert \big)^{2}},
\end{equation}
where $C = C(b,d)$ is a constant depending on $b$ and $d$.
\end{lem}

\hspace{-0.54cm} \begin{prof}
It suffices to show that \eqref{eq0,21} holds for $\lambda$ in each of the four sectors defined by \eqref{eq0,17}-\eqref{eq0,20}. For further use in this proof, let us recall that the relation $\simeq$ is defined by \eqref{eq3.161}.
\\

$\bullet$ For $\big\lbrace \lambda \in \mathbb{C} : \textup{Re} \hspace{0.5mm} \lambda < \mu \big\rbrace$, we have $\textup{dist} \big(\lambda,I\big) = \vert \lambda - \Lambda_{0} \vert$ and by \eqref{eq0,17} $\textup{dist} \big(\varphi_{\mu}(\lambda),\varphi_{\mu}(I)\big) = \vert \varphi_{\mu}(\lambda) \vert = \frac{1}{\vert \lambda - \mu \vert}$. Then $$\frac{\textup{dist} \big(\varphi_{\mu}(\lambda),\varphi_{\mu}(I)\big)}{\textup{dist} \big(\lambda,I\big)} = \frac{1}{\vert \lambda - \mu \vert \vert \lambda - \Lambda_{0} \vert}.$$ So \eqref{eq0,21} holds since $\vert \lambda - \Lambda_{0} \vert \leq C \big(1 + \vert \lambda \vert\big)$ and $\vert \lambda - \mu \vert \leq C \big( 1 + \Vert V \Vert_{\infty} \big) \big(1 + \vert \lambda \vert\big)$.
\\

$\bullet$ For $\lambda \in \left\lbrace \left\vert \lambda - \frac{\mu+\Lambda_{0}}{2} \right\vert \leq \frac{\Lambda_{0}-\mu}{2} \right\rbrace$, we have $\textup{dist} \big(\lambda,I\big) = \vert \lambda - \Lambda_{0} \vert$ and by \eqref{eq0,18} 
\begin{align*}
\textup{dist} \big(\varphi_{\mu}(\lambda),\varphi_{\mu}(I)\big) & = \Big\vert \varphi_{\mu}(\lambda) - \frac{1}{\Lambda_{0} - \mu} \Big\vert \\
& = \frac{\textup{dist} \big(\lambda,I\big)}{\vert \lambda - \mu \vert \vert \Lambda_{0} - \mu \vert}.
\end{align*}
Then \eqref{eq0,21} holds since as above $\vert \lambda - \mu \vert \leq C \big( 1 + \Vert V \Vert_{\infty} \big) \big(1 + \vert \lambda \vert\big)$ and $\vert \Lambda_{0} - \mu \vert \leq C \big( 1 + \Vert V \Vert_{\infty} \big)$.
\\

$\bullet$ For $\lambda \in \big\lbrace \mu \leq \textup{Re} \hspace{0.5mm} \lambda \leq \Lambda_{0} \big\rbrace \setminus \left\lbrace \left\vert \lambda - \frac{\mu+\Lambda_{0}}{2} \right\vert \leq \frac{\Lambda_{0}-\mu}{2} \right\rbrace$, we have $\textup{dist} \big(\lambda,I\big) = \vert \lambda - \Lambda_{0} \vert$ and by \eqref{eq0,18}-\eqref{eq0,19} $$\textup{dist} \big(\varphi_{\mu}(\lambda),\varphi_{\mu}(I)\big) = \vert \textup{Im} \hspace{0.5mm} \varphi_{\mu}(\lambda) \vert = \frac{\vert \textup{Im} \hspace{0.5mm} \lambda \vert}{\vert \lambda - \mu \vert^{2}}.$$ 

$\ast$ For $\lambda$ close to $\mu$ in this domain, $\textup{dist} \big(\lambda,I\big) \simeq \text{constant}$ and $\vert \textup{Im} \hspace{0.5mm} \lambda \vert \simeq \vert \lambda - \mu \vert$ so that $\textup{dist} \big(\varphi_{\mu}(\lambda),\varphi_{\mu}(I)\big) \simeq \frac{1}{\vert \lambda - \mu \vert}$. Then \eqref{eq0,21} holds as above. 

$\ast$ For $\lambda$ close to $\Lambda_{0}$, $\textup{dist} \big(\lambda,I\big) = \vert \lambda - \Lambda_{0} \vert$, $\vert \textup{Im} \hspace{0.5mm} \lambda \vert \simeq \vert \lambda - \Lambda_{0} \vert$ and $\vert \lambda - \mu \vert^{2} \simeq \text{constant}$ so that $\textup{dist} \big(\varphi_{\mu}(\lambda),\varphi_{\mu}(I)\big) \simeq \vert \lambda - \Lambda_{0} \vert$. Then \eqref{eq0,21} holds.

$\ast$ When $\vert \lambda \vert \rightarrow +\infty$, $\textup{Im} \hspace{0.5mm} \lambda \simeq \vert \lambda - \Lambda_{0} \vert$ so that $\textup{dist} \big(\varphi_{\mu}(\lambda),\varphi_{\mu}(I)\big) \simeq \frac{\vert \lambda - \Lambda_{0} \vert}{\vert \lambda - \mu \vert^{2}}$. Then \eqref{eq0,21} holds as above.
\\

$\bullet$ For $\lambda \in \big\lbrace \lambda \in \mathbb{C} : \textup{Re} \hspace{0.5mm} \lambda \geq \Lambda_{0} \big\rbrace$, we have $\textup{dist} \big(\lambda,I\big) = \vert \textup{Im} \hspace{0.5mm} \lambda \vert$ and by \eqref{eq0,20} $$\textup{dist} \big(\varphi_{\mu}(\lambda),\varphi_{\mu}(I)\big) = \vert \textup{Im} \hspace{0.5mm} \varphi_{\mu}(\lambda) \vert = \frac{\vert \textup{Im} \hspace{0.5mm} \lambda \vert}{\vert \lambda - \mu \vert^{2}} = \frac{\textup{dist} \big(\lambda,I\big)}{\vert \lambda - \mu \vert^{2}}.$$ Then \eqref{eq0,21} holds as above, and the Lemma \ref{lem03} is proved. 
\end{prof}

Now for futher use, let us introduce some notations. For $\Lambda_{j} \in E \subset I$, define $r_{j} = \textup{dist} \big(\Lambda_{j},E \setminus \lbrace \Lambda_{j} \rbrace \big)$, $A =\bigcup_{j} B\big(\Lambda_{j},2r_{j}\big)$ and $D = \mathbb{C} \setminus A$ (see the figure below). Corresponding notations on the plane of $z = \varphi_{\mu}(\lambda)$ are $\mathcal{A}$ and $\mathcal{D}$. This means for $\omega \in \varphi_{\mu}(E) = \left\lbrace \frac{1}{\Lambda_{j}-\mu} \right\rbrace_{j}$, $r_{\omega} = \textup{dist} \big(\omega,\varphi_{\mu}(E) \setminus \lbrace \omega \rbrace \big)$, $\mathcal{A} =\bigcup_{\omega} B\big(\omega,2r_{\omega}\big)$ and $\mathcal{D} = \mathbb{C} \setminus \mathcal{A}$.

\begin{center}
\tikzstyle{ddEncadre}=[densely dotted]
\tikzstyle{grisEncadre}=[dashed]
\tikzstyle{grissEncadre}=[fill=gray!60]
\begin{tikzpicture}[scale = 0.8]

\draw [grissEncadre] (4,-1.53) circle (2);
\node at (4,-1.53) {\tiny{$\bullet$}};
\node at (4,-1.85) {\tiny{$\Lambda_{j}$}}; 

\draw [very thick] (-0.5,-1.5) -- (8.5,-1.5);
\node at (9.6,-1.5) {\tiny{$I_{\lambda} = [\Lambda_{0},+\infty)$}};

\node at (-0.5,-1.53) {\tiny{$\bullet$}};
\node at (-0.5,-1.85) {\tiny{$\Lambda_{0}$}};

\node at (4.5,-1.2) {\tiny{$r_{j}$}};
\draw [<->] (4,-1.35) -- (5,-1.35);

\node at (3.5,-1.2) {\tiny{$r_{j}$}};
\draw [<->] (3,-1.35) -- (4,-1.35);

\draw (3,-1.53) circle (2);
\node at (3,-1.53) {\tiny{$\bullet$}};
\node at (3,-1.85) {\tiny{$\Lambda_{j-1}$}};

\draw (5,-1.53) circle (2);
\node at (5,-1.53) {\tiny{$\bullet$}};
\node at (5,-1.85) {\tiny{$\Lambda_{j+1}$}};

\draw (6,-1.53) circle (2);
\node at (6,-1.53) {\tiny{$\bullet$}};
\node at (6,-1.85) {\tiny{$\Lambda_{j+2}$}};

\draw [<-] (5.3,-0.8) -- (8.5,0.5);
\node at (8.9,0.7) {\tiny{$B\big(\Lambda_{j},2r_{j}\big)$}};

\end{tikzpicture}

\vspace*{0.2cm}

\textsc{Figure $5$.} Sets $I_{\lambda} = [\Lambda_{0},+\infty)$ and $A =\bigcup_{j} B\big(\Lambda_{j},2r_{j}\big)$.
\end{center}
Note that by \eqref{eq1,00}, we have $r_{j} = 2b$, $A = \bigcup_{j} B\big(\Lambda_{j},4b\big)$ and up to constant factor $\varphi_{\mu}(A) = \mathcal{A}$.
We have the following lemma.

\begin{lem}\label{lem04}
With the notations above, the following estimates hold.

\textup{(i)} For any $\lambda \in D$ and any $\varphi_{\mu}(\lambda) \in \mathcal{D}$,
\begin{equation}\label{eq0,22}
\begin{split}
\small{\frac{\textup{dist} \big(\lambda,E\big)}{2}} & \small{\leq \textup{dist} \big(\lambda,I\big) \leq \textup{dist} \big(\lambda,E\big)}. \\
\small{\frac{\textup{dist} \big(\varphi_{\mu}(\lambda),\varphi_{\mu}(E)\big)}{2}} & \small{\leq \textup{dist} \big(\varphi_{\mu}(\lambda),\varphi_{\mu}(I)\big) \leq \textup{dist} \big(\varphi_{\mu}(\lambda),\varphi_{\mu}(E)\big)}.
\end{split}
\end{equation}

\textup{(ii)} For any $\lambda \in A$,
\begin{equation}\label{eq0,23}
\small{\textup{dist} \big(\varphi_{\mu}(\lambda),\varphi_{\mu}(E)\big) \geq \frac{C \hspace{0.08cm} \textup{dist} \big(\lambda,E\big)}{\big( 1 + \Vert V \Vert_{\infty} \big)^{2} \big( 1 + \vert \lambda \vert \big)^{2}}}.
\end{equation}
\end{lem}

\hspace{-0.54cm} \begin{prof}
(i): We prove only the first estimate in \eqref{eq0,22}. The same holds for the second. Obviously $\textup{dist} \big(\lambda,I\big) \leq \textup{dist} \big(\lambda,E\big)$. For $\lambda \in D$, let $\Lambda_{j} \in E$ such that 
\begin{equation}\label{eq0,24}
\small{\textup{dist} \big(\lambda,E\big) = \vert \lambda - \Lambda_{j} \vert \geq 2r_{j} \quad \left( \Longleftrightarrow \frac{\textup{dist} \big(\lambda,E\big)}{2} \geq r_{j} \right)}.
\end{equation}
Since $\textup{dist} \big(\lambda,E\big) \leq \textup{dist} \big(\lambda,I\big) + r_{j}$, $i.e.$ $\textup{dist} \big(\lambda,E\big) - r_{j} \leq \textup{dist} \big(\lambda,I\big)$, then \eqref{eq0,24} implies that $\textup{dist} \big(\lambda,E\big) - \frac{\textup{dist} \big(\lambda,E\big)}{2} \leq \textup{dist} \big(\lambda,I\big)$. That means $$\small{\frac{\textup{dist} \big(\lambda,E\big)}{2} \leq \textup{dist} \big(\lambda,I\big)}.$$
(ii): Obviously all points $\lambda \in A$ are of the form $\vert \textup{Im} \hspace{0.5mm} \lambda \vert \leq 4b$ and $\Lambda_{j} \leq \textup{Re} \hspace{0.5mm} \lambda \leq \Lambda_{j+1}$ for some $\Lambda_{j} \in E$ such that $\textup{dist} \big(\lambda,E\big) = \vert \lambda - \Lambda_{j} \vert$ \big(or $\textup{dist} \big(\lambda,E\big) = \vert \lambda - \Lambda_{j+1} \vert$\big). For $\textup{dist} \big(\lambda,E\big) = \vert \lambda - \Lambda_{j} \vert$, we have 
$$\small{\textup{dist} \big(\varphi_{\mu}(\lambda),\varphi_{\mu}(E)\big) = \Big\vert \frac{1}{\lambda - \mu} - \frac{1}{\Lambda_{j} - \mu} \Big\vert = \frac{\textup{dist} \big(\lambda,E\big)}{\vert \lambda - \mu \vert \vert \Lambda_{j} - \mu \vert}}.$$ 
Then \eqref{eq0,23} holds since $\vert \lambda - \mu \vert \leq C \big( 1 + \Vert V \Vert_{\infty} \big) \big(1 + \vert \lambda \vert\big)$ and $\vert \Lambda_{j} - \mu \vert \leq 2 \vert \lambda - \mu \vert$ for $\textup{Re} \hspace{0.5mm} \lambda > \Lambda_{0}$. This completes the proof.
\end{prof}

Now we turn back to the proof of Theorem \ref{theo3}. Lemma \ref{lem02} together with \eqref{eq0,15} and \eqref{eq0,151} show that
$$\sum_{z \in \sigma_{d}(H_{\perp})} \frac{\textup{dist} \big(\lambda,E\big)^{p}}{\big( 1 + \vert \lambda \vert \big)^{2p}} \leq C_{1} \Vert F \Vert_{L^{p}}^{p} \big( 1 + \Vert V \Vert_{\infty} \big)^{2p},$$
where $C_{1} = C(p,b,d)$ is a constant depending on $p$, $b$ and $d$. This completes the proof of Theorem 2.2.

\textit{Acknowledgments}. This work is partially supported by ANR NOSEVOL-11-BS01-019 01. The author is grateful to V. Bruneau and S. Kupin for valuable discussions on the subject of this article.

\end{document}